\documentclass[%
 reprint,
 amsmath,amssymb,
 aps,
]{revtex4-2}

\usepackage{url}
\usepackage[%
  colorlinks=true,
  urlcolor=blue,
  linkcolor=blue,
  citecolor=blue
]{hyperref}

\usepackage{placeins} 
\usepackage{xcolor}
\usepackage{xparse}
\usepackage{cleveref}

\usepackage{graphicx}
\usepackage{bm}

\begin{document}

\title{Antivax movement and epidemic spreading in the era of social networks:  nonmonotonic effects, bistability and network segregation}

\author{Marcelo A. Pires$^{1}$}
\thanks{piresma@cbpf.br}

\author{Andre L. Oestereich$^{2}$}
\thanks{andreoestereich@uceff.edu.br}

\author{Nuno Crokidakis$^{3}$}
\thanks{nuno@mail.if.uff.br}

\author{S\'{\i}lvio M. \surname{Duarte~Queir\'{o}s}$^{1,4,5}$}
\thanks{sdqueiro@cbpf.br}

\affiliation{
$^{1}$Centro Brasileiro de Pesquisas F\'isicas, Rio de Janeiro/RJ, Brazil
\\
$^{2}$Unidade Central De Educação Faem Faculdade, Itapiranga/SC, Brazil
\\
$^{3}$Instituto de F\'isica, Universidade Federal Fluminense, Niter\'oi/RJ, Brazil
\\
$^{4}$National Institute of Science and Technology for Complex Systems, Brazil
\\
$^{5}$i3N, Universidade de Aveiro, Campus de Santiago, 3810-193 Aveiro, Portugal
}


\date{\today}

\begin{abstract}

In this work, we address a  multicoupled dynamics on complex networks
 with tunable structural segregation.
Specifically, we work on a networked epidemic spreading under a vaccination campaign with agents in favor and against the vaccine. Our results show that such coupled dynamics exhibits a myriad of phenomena such as nonequilibrium transitions accompanied by bistability. Besides we observe the emergence of an intermediate optimal segregation level where the community structure enhances negative opinions over vaccination but counterintuitively hinders - rather than favoring - the global disease spreading. Thus, our results hint vaccination campaigns  should avoid policies that end up segregating excessively anti-vaccine groups so that they effectively work as echo chambers in which individuals look to confirmation without jeopardising the safety of the whole population.


\end{abstract}

\maketitle



\section{\label{sec:intro}Introduction}

In such a complex contemporary society where elements -- people and events -- influence one another and feedback at different scales~\cite{parisi1999}, the application of tools set forth in Statistical Physics in order to cope with collective phenomena has gained prominence in other areas such as Biology and Medicine, Social Sciences, and Humanities, which have put quantitative tools in the methodologies they apply~\cite{thurner2018,stauffer2006biology,castellano2009statistical,de2013evolution,galam2008sociophysics,sen2014sociophysics}. 

 Accordingly, phenomena in which there is a change in the collective behavior displayed by a social system have turn into an appropriate field for the application of such techniques~\cite{pastor2015,arruda2018,wang2017}; among the several different instances we can find important contributions within the spreading of epidemics as well as opinion dynamics (see eg~\cite{wang2019coevolution} and Sec.~\ref{sec:litrev}).
In spite of the fact that the two subject-matters are not related at first, the dissemination of a causal relationship between neurological disorders and vaccinations~\cite{gasparini2015} has prompted an urban myth that ultimately has jeopardized the elimination of the disease in countries with a very high Human Development Index as the USA~\cite{feemster2020}.

Next the manuscript is organized as follows: in Sec.~\ref{sec:litrev}, we establish the state-of-the-art of the problem; in Sec.~\ref{sec:level1}, we introduce our model for the combined dynamics of opinion and contagion; in Sec.~\ref{sec:resudisc}, we discuss the results for the model; and in Sec.~\ref{sec:final}, we present our final observations on the work and future perspectives about it.


\section{\label{sec:litrev}Literature review}
Modular networks~\cite{2014nematzadehFFA}
are generated by an algorithm that leads to networks with an architecture of communities. A given node in each community can be connected to nodes of the same community (intracommunity links) and/or to nodes of the other community (intercommunity links).

The impact of the network modularity in spreading processes has been investigated in recent years. Since the results introduced in Ref.~\cite{2014nematzadehFFA}, a series of works were published regarding the subject of optimal network modularity; therein, the authors showed that modular structure may have counterintuitive effects on information diffusion. Indeed, it was discussed that the presence of strong communities in modular networks can facilitate global diffusion by improving local intracommunity spreading.

Still in relation to modular networks, it was recently found that an optimal community structure that maximizes spreading dynamics which can pave the way to rich phase diagrams with exhibiting first-order phase transitions~\cite{su2018optimal}. Within the same context, the authors in Ref.~\cite{wu2016optimal} discussed the impact of social reinforcement in information diffusion. They also found optimal multi-community network modularity for information diffusion, i.e., depending on the range of the parameters the multi-community structure can facilitate information diffusion instead of hindering it.

Regarding biological systems, it was recently found there is a nonlinear relation between modularity and global efficiency in animal networks, with the latter peaking at intermediate values of the former~\cite{romano2018social}. In addition, in neural networks there exists an optimal modularity for memory performance, where a balance between local cohesion and global connectivity is established, allowing optimally modular networks to remember longer~\cite{rodriguez2019optimal}.

 The authors in Ref.~\cite{cui2018close} studied the importance of close and ordinary social contacts in promoting large-scale contagion and found an optimal fraction of ordinary contacts for outbreaks at a global scale. With respect to correlations in complex networks, it was found that constraining the mean degree and the fraction of initially informed nodes, the optimal structure can be assortative (modular), core-periphery or even disassortative \cite{curato2016optimal}. Other recent works leading with optimal modularity in networks can be found in \cite{nematzadeh2018optimal,PhysRevE.102.052316}.

In a recent work \cite{valdez2020epidemic}, it was proposed a model of disease spreading in a structural modular complex network and studied how the number of bridge nodes $n$ that connect communities affects disease spreading. It was verified that near the critical point as $n$ increases, the disease reaches most of the communities, but each community has only a small fraction of recovered nodes. Moreover, a combination of social networks with game theory was studied in Refs.~\cite{Nowak2011,Feng2019}.

Disease information can spark strong emotions like fear --- or even panic --- that would affect behavior during an epidemic. The authors in \cite{bi2019modeling} considered an agent-based model that assumes that agents can obtain a complete picture of the epidemic via information from local daily contacts or global news coverage. Those results helped conclude that such model can be used to mimic real-world epidemic situations and explain disease transmission, behavior changes, and distribution of prevalence panic. Game theory was also considered to reproduce the decision-making process of individuals during the evolution of a disease. In \cite{zhao2018risk} a spatial evolutionary game was coupled to a SIR model, and the results showed that protective behaviors decrease the numbers of infected individuals and delay the peak time of infection. The study also concluded that increased numbers of risk-averse individuals and preemptive actions can more effectively mitigate disease transmission; however, changes in human behavior require a high social cost (such as avoidance of crowded places leading to absences in schools, workplaces, or other public places).

A recent work considered a coupled behavior-change and infection in a structured population characterized by homophily and outgroup aversion~\cite{smaldino2020coupled}. It was found that homophily can either increase or decrease the final size of the epidemic depending on its relative strength in the two groups. In addition, homophily and outgroup aversion can also produce a `second wave' in the first group that follows the peak of the epidemic in the second group.

Models of opinion dynamics were applied in the context of  opinions about vaccination  (pro versus anti-vaccine) without coupling an epidemic process~\cite{galam2010public}. Later, kinetic opinion dynamics were  coupled to classical epidemic models in order to study the feedback among risk perception, opinions about vaccination, and the disease spreading. In \cite{pires2017dynamics} it was found that the engagement of the pro-vaccine individuals can be crucial for stopping the epidemic spreading. On the other hand, the work \cite{2018piresOC} found counterintuitive outcomes like the fact that an increment in the initial fraction of the population that is pro-vaccine can lead to smaller epidemic outbreaks in the short term, but it also contributes to the survival of the chain of infections in the long term.

Recently, the anti-vaccine sentiment was treated as a cultural pathogen. The authors in \cite{mehta2020modelling} modeled it as a 'infection' dynamics. The authors showed that interventions to increase vaccination can potentially target any of three types of transitions - decreasing sentiment transmission to undecided individuals, increasing pro-vaccine decisions among undecided individuals, or increasing sentiment switching among anti-vaccine individuals.

We previously cited anti-vaccine opinions, thus it is important to mention some recent discussion about the global anti-vaccine movement. Since the online discussions dominate the social interactions in our modern world, the propagation of such anti-vaccine opinions is growing fast. A recent report  noted  that  31  million  people  follow  anti-vaccine  groups  on  Facebook,  with  17  million  people  subscribing    to    similar    accounts    on  YouTube \cite{burki2020online}. The authors in \cite{johnson2020online} recently pointed that if the current trends continue, anti-vaccine views will dominate online discussion in 10 years. The importance of anti-vaccine movement is fundamental for the evolution of COVID-19 outbreak. Indeed, the authors in \cite{buonomo2020effects} called attention to the fact that it is a key point to qualitatively assess how the administration of a vaccine could affect the COVID-19 outbreak, taking into account of the behavioral changes of individuals in response to the information available on the status of the disease in the community. According to a study published in August 2020, nearly one in four adults would not get a vaccine for COVID-19 \cite{boyon} and in some countries, more than half of the population would not get it, including Poland and France \cite{curielantivax}. In September 2020, it was verified that only 42 percent of	Americans said \textit{yes}	to receiving a future COVID-19 vaccine, across	all	political sides. It	means that even	in a best-case scenario where a future	high performing	vaccine	is	95$\%$	effective in an	individual,	it would only impact 42x95$\approx 40\%$	of	the
population, which is way below predicted thresholds	for	herd immunity \cite{johnsonnotsure}.

\begin{figure}[h]
    \centering
    \includegraphics[width=0.29\textwidth]{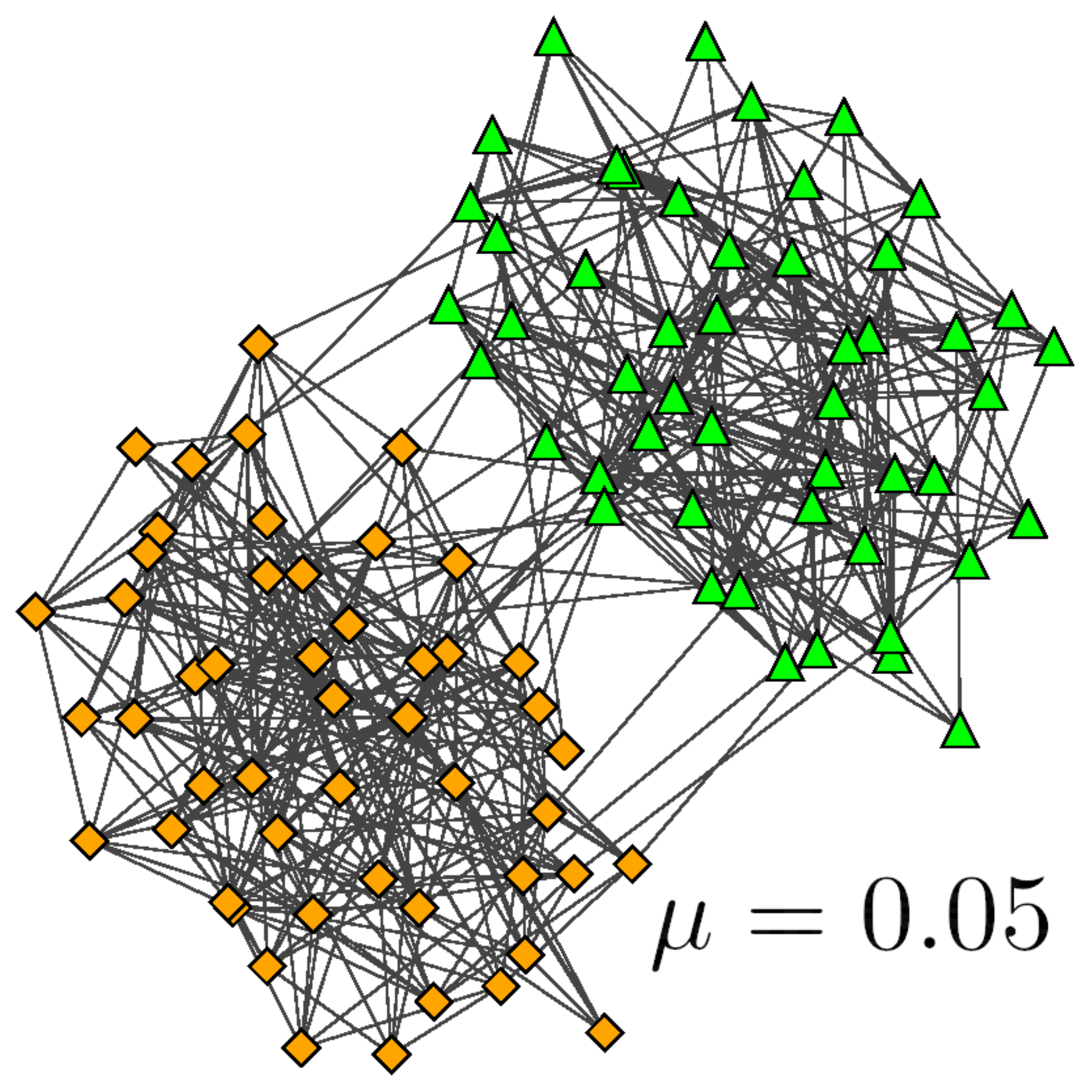}
    \includegraphics[width=0.29\textwidth]{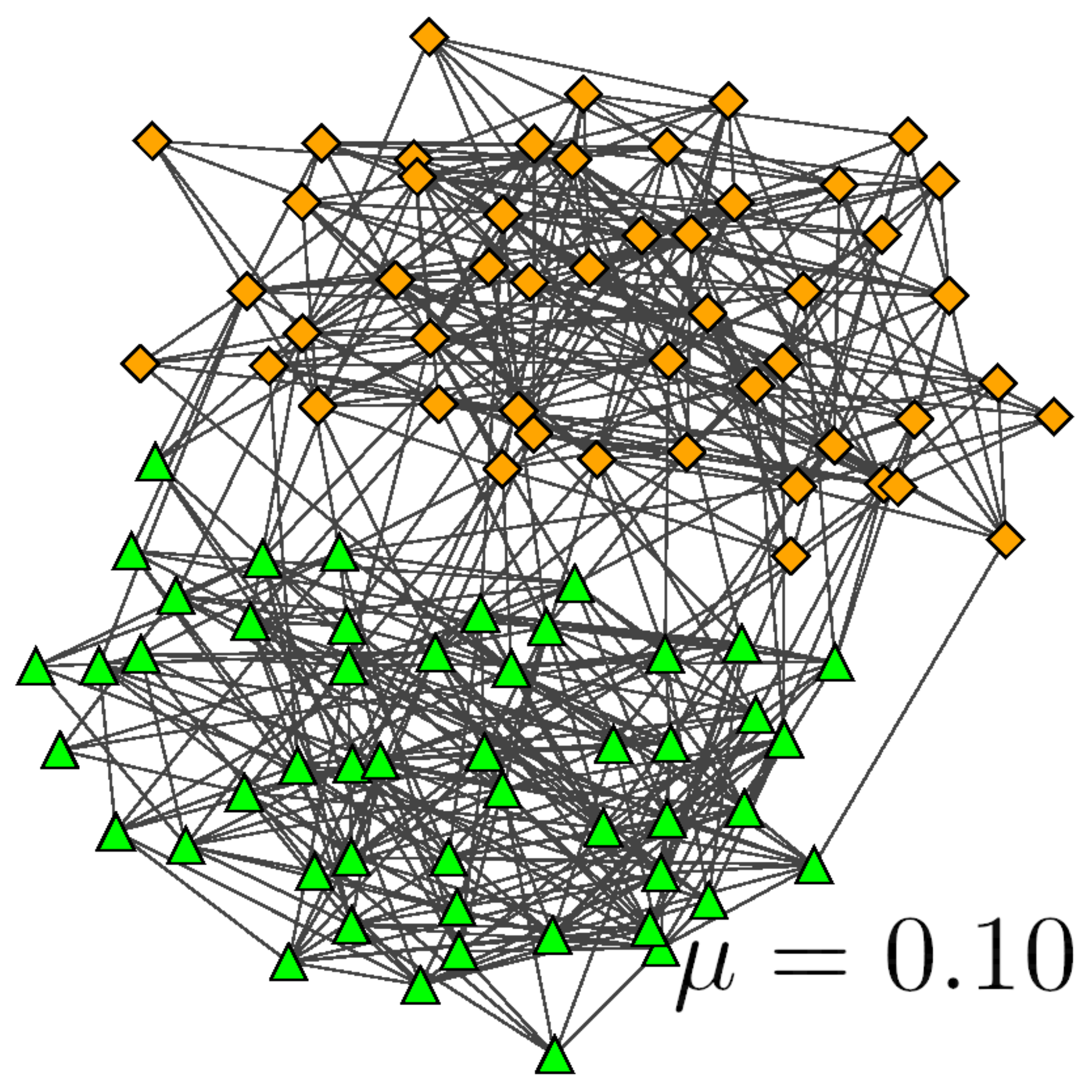}
    \includegraphics[width=0.29\textwidth]{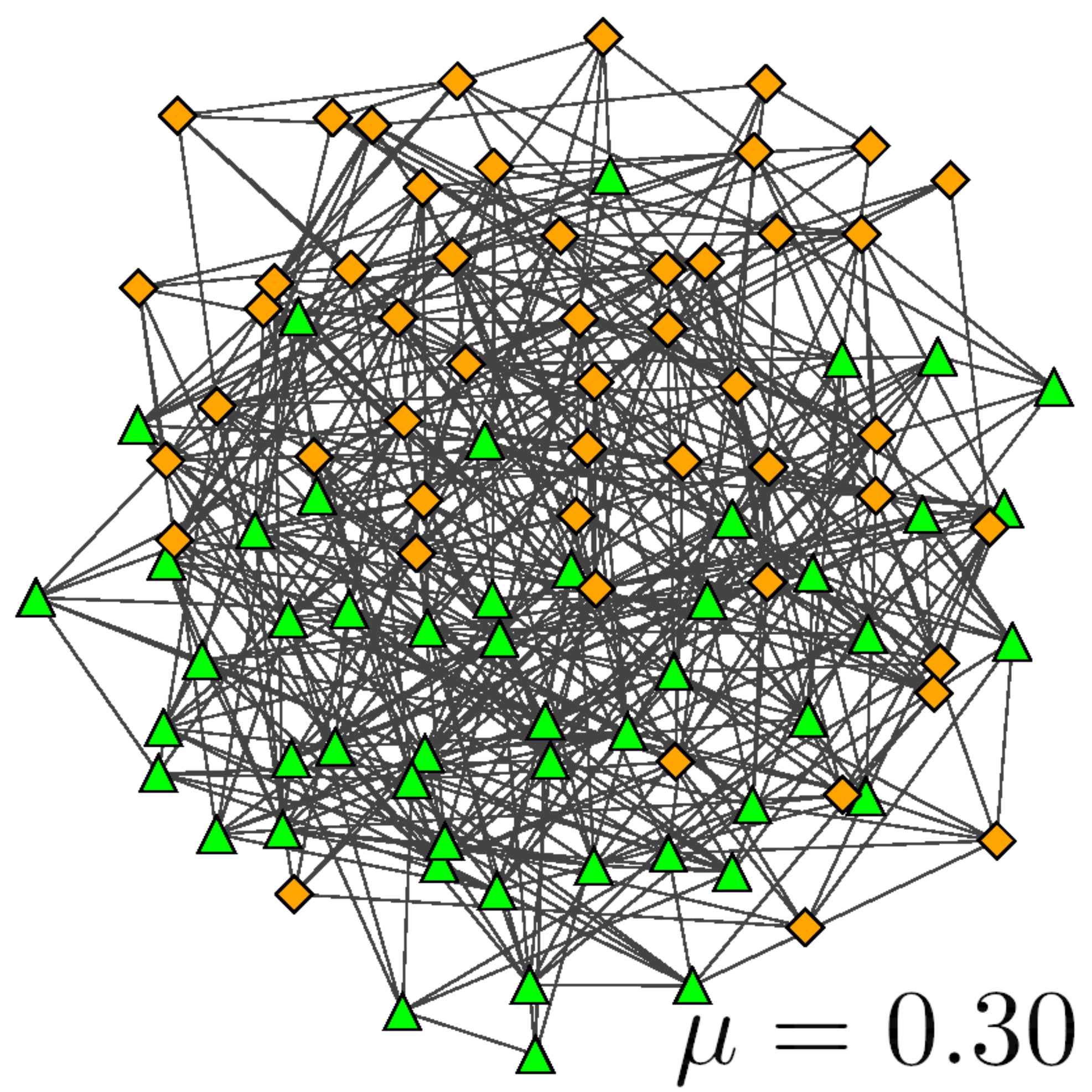}
    \caption{Examples of modular networks with $N=100$, $\langle k \rangle = 10$ for different values of $\mu$. The parameter $\mu$ is the community interconnectivity:   small values of $\mu$  means few intercommunities bridges which implies strong community structure, ie strong modularity/segregation. In these examples we can see the strengthening of the community structure for lower values of $\mu$.}
\end{figure}

\begin{figure}[h]
    \centering
    \includegraphics[width=0.49\textwidth]{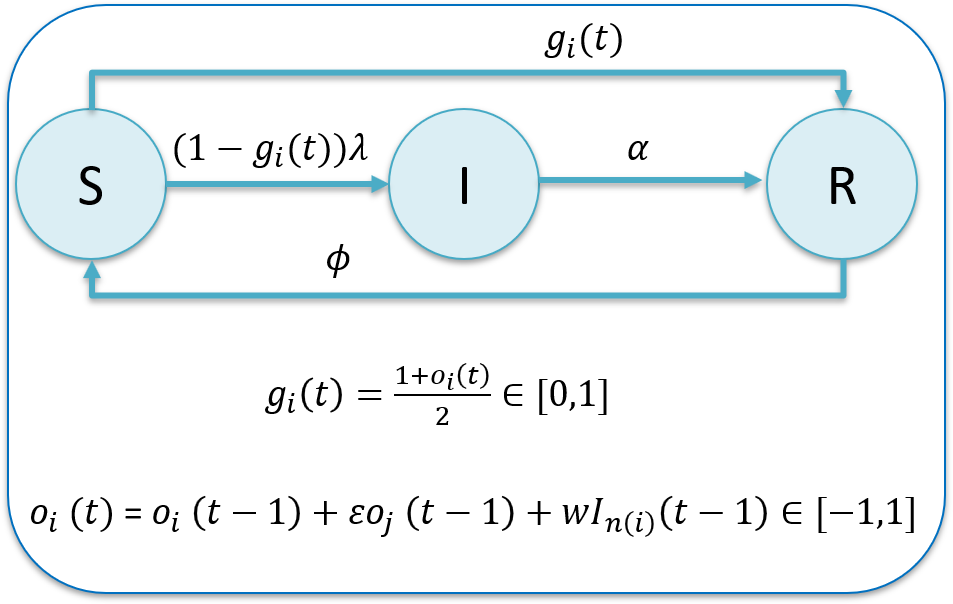}
    \caption{Coupled vaccination and continuous opinion dynamics.}
    \label{fig:fullmodel}
\end{figure}

\begin{figure*}[t]
\centering
\includegraphics[width=0.85\textwidth]{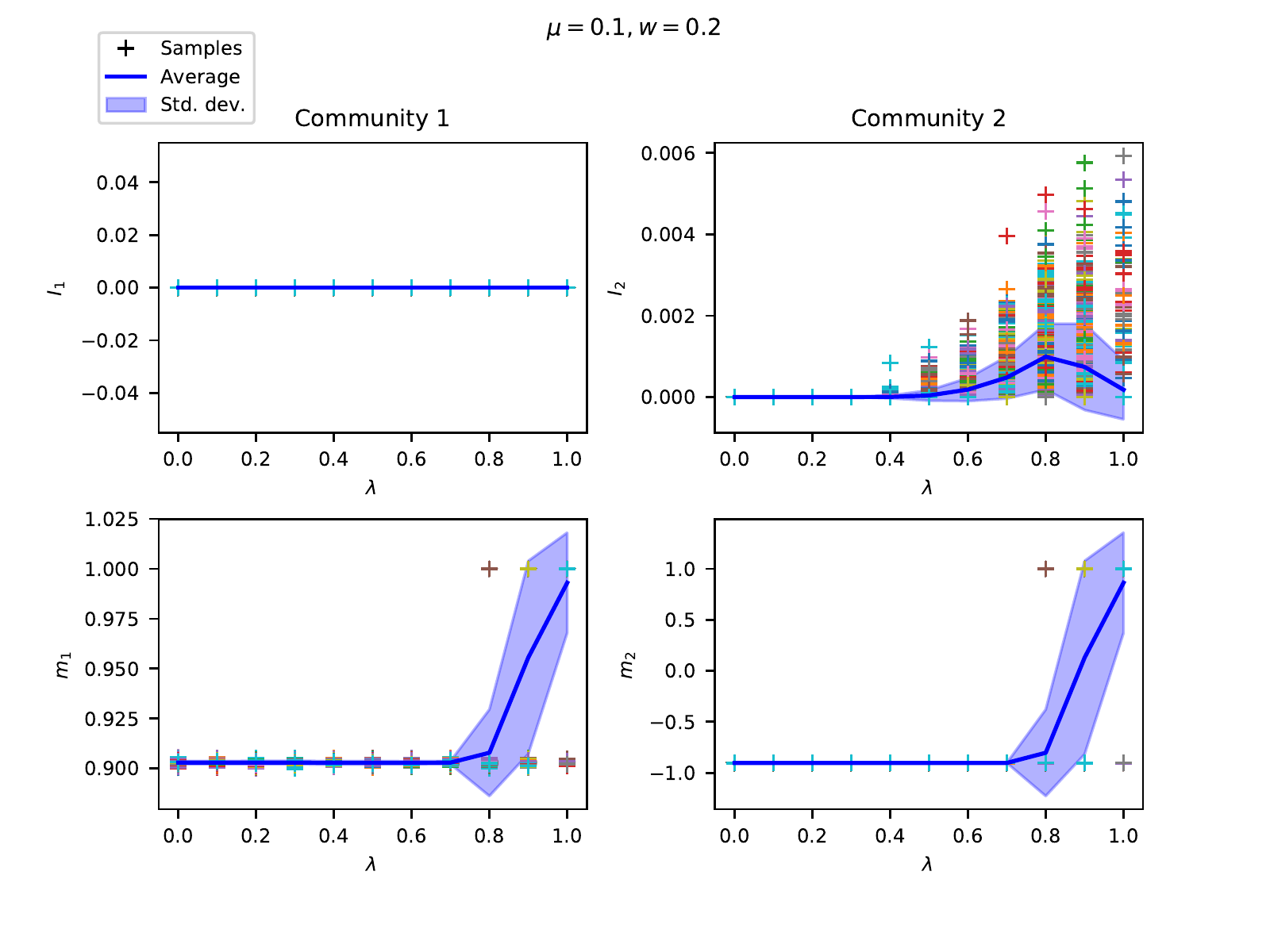}
\caption{ Steady-state for the spreading measure $I_i$ and collective opinion $m_i$ for each community $i=\{1,2\}$. Symbols are the steady-state outcome for each sample, i.e., each symbol is the result from each Monte Carlo realization. Results for $\mu=0.1$. For this high level of segregation, each community ends up preserving the sign of its initial opinion. Besides, the chain of contagion starts in the community $2$ and cannot become permanent in the community $1$.
}
\label{fig:steadystate-1}
\end{figure*}

\begin{figure*}[t]
\centering
\includegraphics[width=0.85\textwidth]{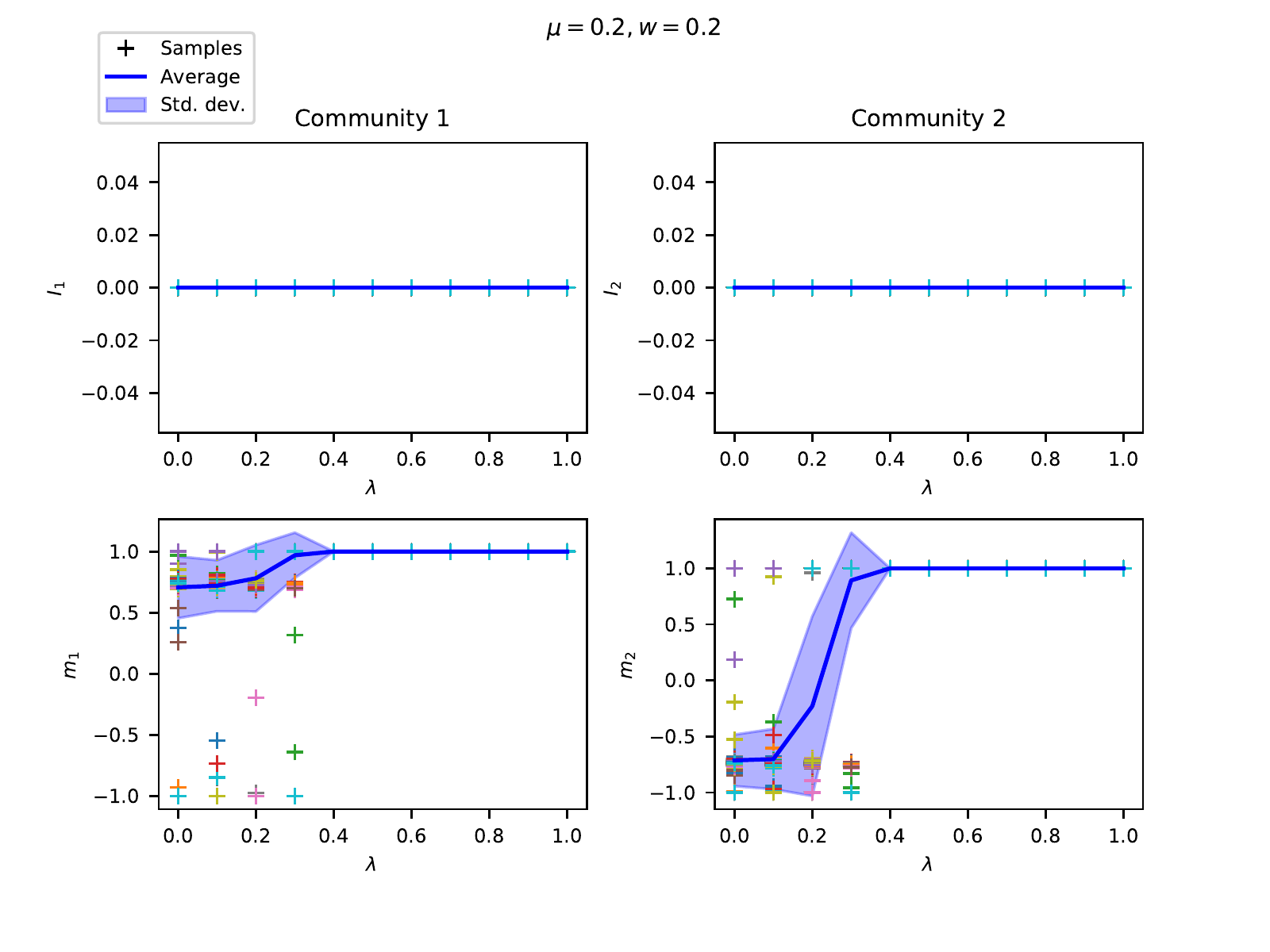}
\caption{Steady-state for the spreading measure $I_i$ and collective opinion $m_i$ for each community $i=\{1,2\}$. Symbols are the steady-state outcome for each sample, ie, each symbol is the result from each Monte Carlo realization. Results for $\mu=0.2$. For this intermediate level of segregation, there is the possibility for a switch of opinion in the community $2$ (seed). The epidemic spreading does not survive at the global and local levels.
}
\label{fig:steadystate-2}
\end{figure*}

\begin{figure*}[t]
\centering
\includegraphics[width=0.85\textwidth]{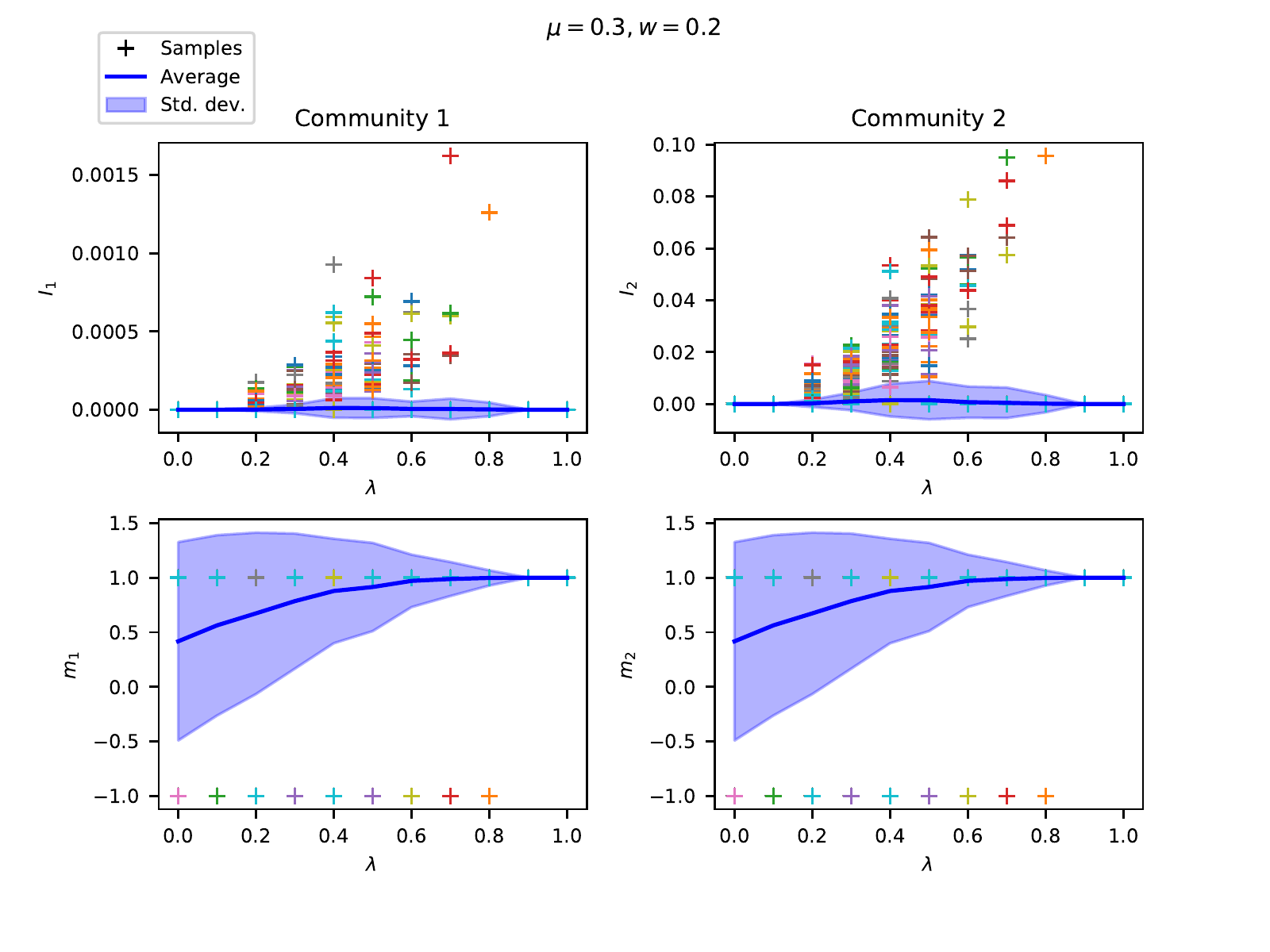}
\caption{Steady-state for the spreading measure $I_i$ and collective opinion $m_i$ for each community $i=\{1,2\}$. Symbols are the steady-state outcome for each sample, i.e., each symbol is the result from each Monte Carlo realization. Results for $\mu=0.3$. For this low level of segregation, there is the possibility for a switch of opinion in both communities. The epidemic dynamics can survive if the contagion is not too aggressive (intermediate values of  $\lambda$) .}
\label{fig:steadystate-3}
\end{figure*}

\begin{figure*}[t]
\centering
\includegraphics[width=0.85\textwidth]{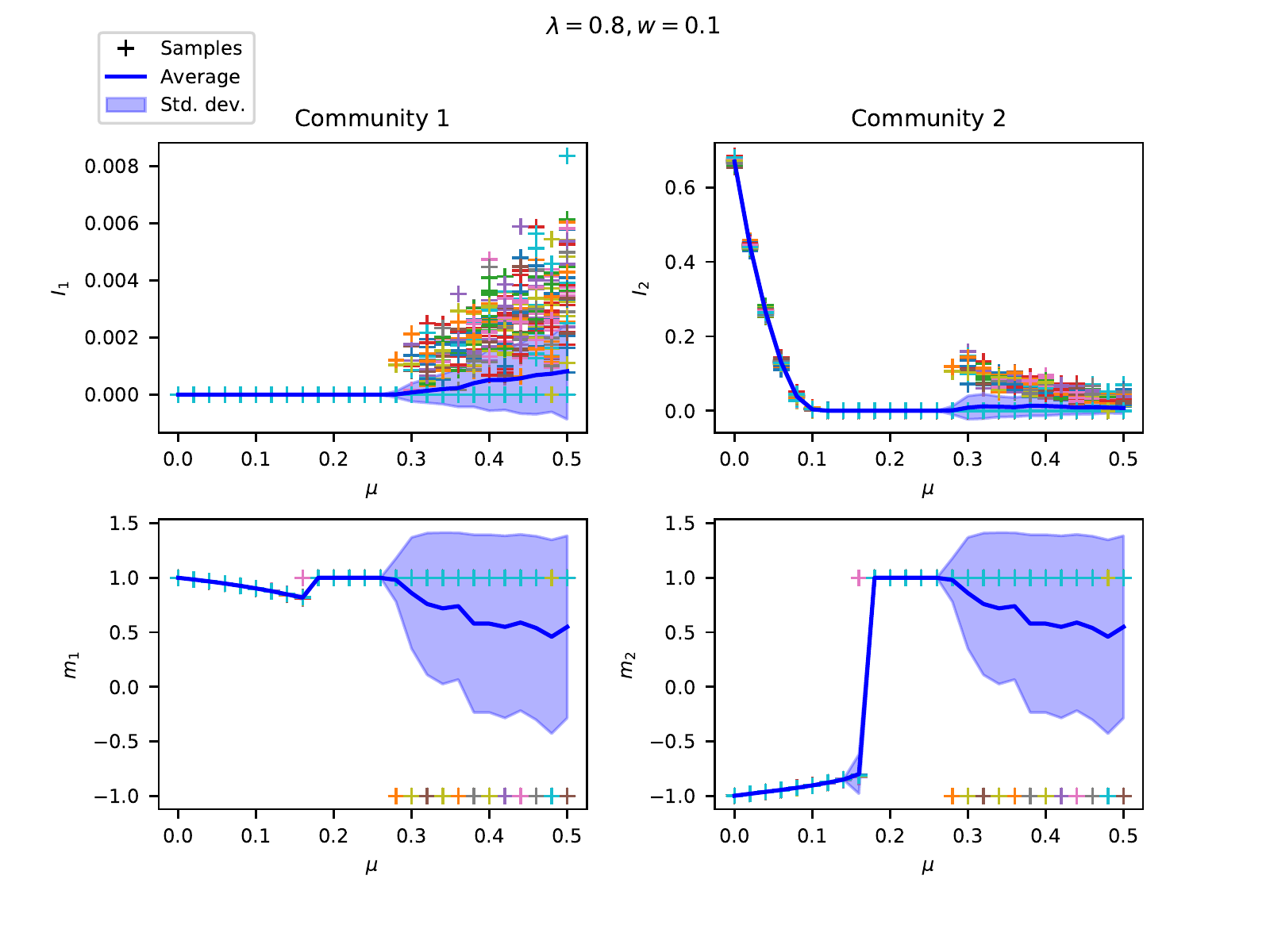}
\caption{Steady-state for the spreading measure $I_i$ and collective opinion $m_i$ for each community $i=\{1,2\}$. Symbols are the steady-state outcome for each sample, i.e., each symbol is the result from each Monte Carlo realization. Results for $w=0.1$ and $\lambda=0.8$.}
\label{fig:steadystate-4}
\end{figure*}

\section{\label{sec:level1}Model}

\subsection{I: Opinion dynamics}

Even though payoff-based models have been employed to address the problem of vaccination dynamics (for instance see \cite{Nowak2011,Feng2019,wang2016statistical} and the references therein), there is an alternative  approach that is based on the coupling of epidemic and psychosocial factors  that have been provided a successful  modelling of phenomena related to vaccination dynamics~\cite{salathe2008effect,wu2016dynamics,coelho2009dynamic,smaldino2020coupled,voinson2015beyond,mehta2020modelling,Feng2017}. In this work,  we follow such second methodology. Specifically, based on Refs.~\cite{2018piresOC},\cite{2010lallouacheCCC} we consider an agent-based   dynamics
in which the opinion about vaccination, $o_i\in [-1,1]$, of each agent, $i$, evolves with
\begin{equation}
      o_i(t+1)=o_i(t)+\epsilon o_j(t)+wI_{n(i)}(t),
      \label{eq:evol}
\end{equation}
%
%
A negative (positive) value of $o_i$ represents an individual $i$ supporting anti-vaccine (pro-vaccine) opinion. Equation~\eqref{eq:evol} takes into account the agent's opinion,  $o_{i}(t+1)$,  depends on multiple factors: (i) his previous opinion $ o_i(t)$; (ii) the peer pressure  exerted by a randomly selected neighbor, $j$, modulated by a stochastic heterogeneity $\epsilon$, uniformly distributed in the interval $[0, 1]$; (iii) the proportion of infected neighbors, $I_{n(i)}(t)$, modulated by a risk perception parameter, $w$. Notice that, in Eq. \eqref{eq:evol}, if the value of the opinion exceeds (falls below) the value $1 (-1)$, then it adopts the extreme value $1 (-1)$ \cite{2010lallouacheCCC}.

The opinion dynamics regarding the vaccination campaign is coupled with the epidemic dynamics, due to the factor $I_{n(i)}(t)$ in Eq. \eqref{eq:evol}.

\subsection{II: epidemics-vaccination  dynamics}

Based on \cite{pires2017dynamics,2018piresOC} (and references therein), we define the transitions among the epidemic compartments as follows:

\begin{itemize}
\item $S \stackrel{g_{i}}{\rightarrow} R$: a Susceptible agent $i$ becomes Vaccinated with probability $g_i$;

\item $S \stackrel{(1-g_{i})\lambda}{\rightarrow} I$:  a Susceptible  agent $i$ becomes Infected with probability $(1-g_i)\lambda$ if he is in contact with an Infected agent;

\item $I \stackrel{\alpha}{\rightarrow} S$: an Infected agent $i$ recovers with probability  $\alpha$;

\item $R \stackrel{\phi}{\rightarrow} S$: a immune agent $i$ becomes Susceptible again with the resusceptibility probability $\phi$. We assume that Vaccinated and Recovered agents are in the same compartment\cite{zeng2005complexity,rao2019complicated,moneim2005threshold,lahrouz2012complete,doutor2016optimal}.

\end{itemize}

The vaccination probability $g_{i}$  of an agent $i$ is proportional to his opinion about vaccination $-1\leq o_i\leq1$:
\begin{align}
g_i(t) = \frac{1+o_{i}(t)}{2} \in [0,1]
      \label{eq:gamma}
\end{align}

Despite the differences, the modeling of the coupling between disease and opinion evolution is still  a open subject. In this work, we consider the two dynamics have the same time scale. An overview of our model is shown in Fig.\ref{fig:fullmodel}. An element in this problem which is still focus of debate concerns the timescale of each dynamics, epidemic and opinion. On the one hand, it is often assumed in the epidemiological literature~\cite{wu2016dynamics,voinson2015beyond,coelho2009dynamic,velasquez2017interacting} that the two timescales are equivalent. At first, this can be understood as a simplification it captures the mass vaccination campaigns governments swiftly implement  in order to avoid disease outbreaks.
On the other hand, it is possible to assume different timescales of evolution of the diseases and opinions about the disease~\cite{ventura2021role,da2019epidemic}.
In this work we consider the first approach of equality between the two timescales.

\subsection{Community structure}

Based on Ref.~\cite{2019oestereichPC} and related literature, we start by picking the first $N_1=N/2$ of the $N$ nodes and attaching them to the community $1$, and assigning the other $N_2=N - N_1$ nodes to community $2$. We then proceed by randomly assigning $(1-\mu)M$ connections among pairs of nodes from the same community and $\mu M$ connections are randomly distributed among pairs of nodes that belong to distinct communities, where $M=N\,k/2$ and $k$ is the network average degree \cite{2019oestereichPC}.

The parameter $\mu$ regulates the community strength:  large values of $\mu$ means more ties between the two communities consequently a weaker community organization. Another way to control the network structure -- especially in the formation of the echo chambers -- is by considering rewiring~\cite{Wang2020}.

\subsection{Initial condition}

We consider that community $1$ holds a positive stance on vaccination, whereas the community $2$ holds a negative opinion about that. We also assume the chain of infections starts in community $2$, because $o_i<0$ leads to a low propensity for the agents to get vaccinated, which is naturally more relevant. If the epidemic started in community $1$, pro-vaccine opinions, $o_i>0$, would induce a higher probability for an agent to get vaccinated that ultimately would end up disrupting the chain of contagions.


Let $U(a,b)$ be a single random value from a uniform distribution in the range $[a,b]$.

At $t=0$, we set:

\begin{itemize}
    \item For i in $0 \ldots N/2-1$: (community 1: $o_i>0$; $0\%$ infected)
    \begin{itemize}
        \item $o_i \sim U(0,1) $
        \item status(i) =  S
    \end{itemize}
    \item For i in $N/2 \ldots N-1$: (community 2: $o_i<0$; $1\%$ of infected)
    \begin{itemize}
        \item $o_i  \sim U(-1,0) $
        \item status(i) =  S with probability $0.99$
        \item status(i) =  I with probability $0.01$
    \end{itemize}

\end{itemize}

\section{\label{sec:resudisc}Results and Discussion}

\begin{figure}[h]
    \centering
    \includegraphics[width=0.49\textwidth]{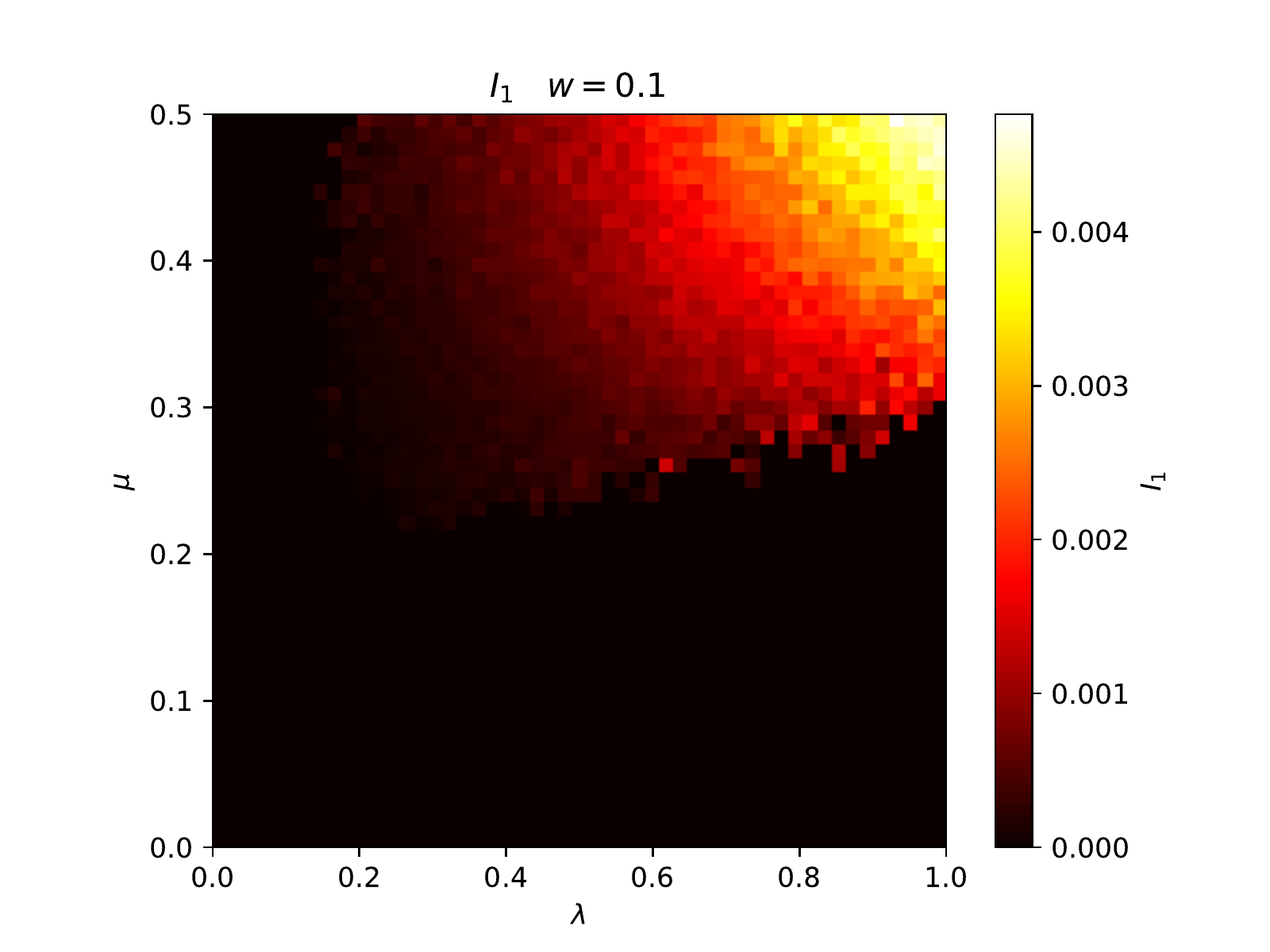}
    \includegraphics[width=0.49\textwidth]{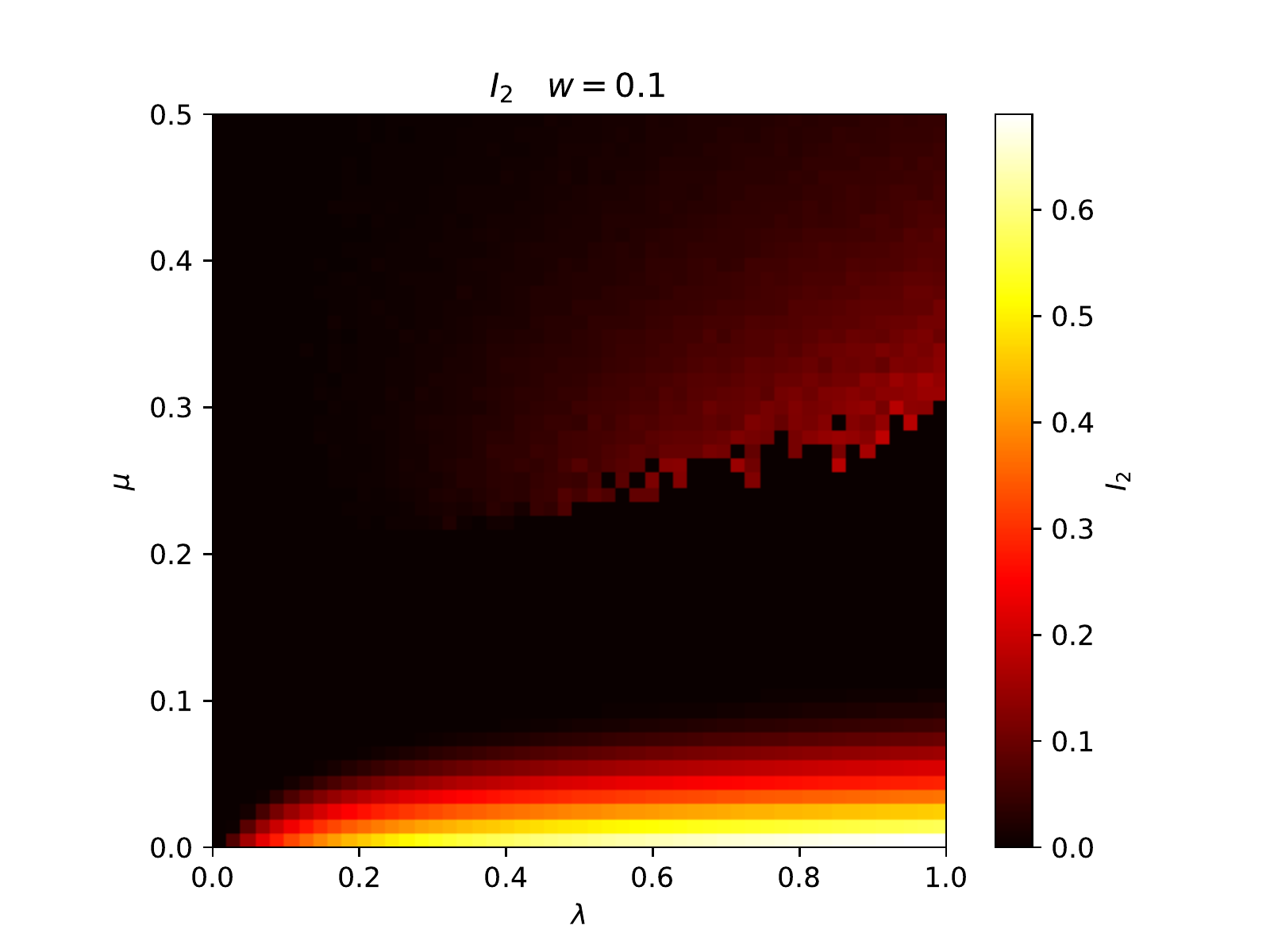}
    \caption{Steady-state for the spreading measure $I_i$ for community 1, above, and 2, bellow. The colors indicate the average of non zero steady-state outcomes of all 200 samples. Results for $w=0.1$.}
    \label{fig:new_results1}
\end{figure}

\begin{figure}[h]
    \centering
    \includegraphics[width=0.49\textwidth]{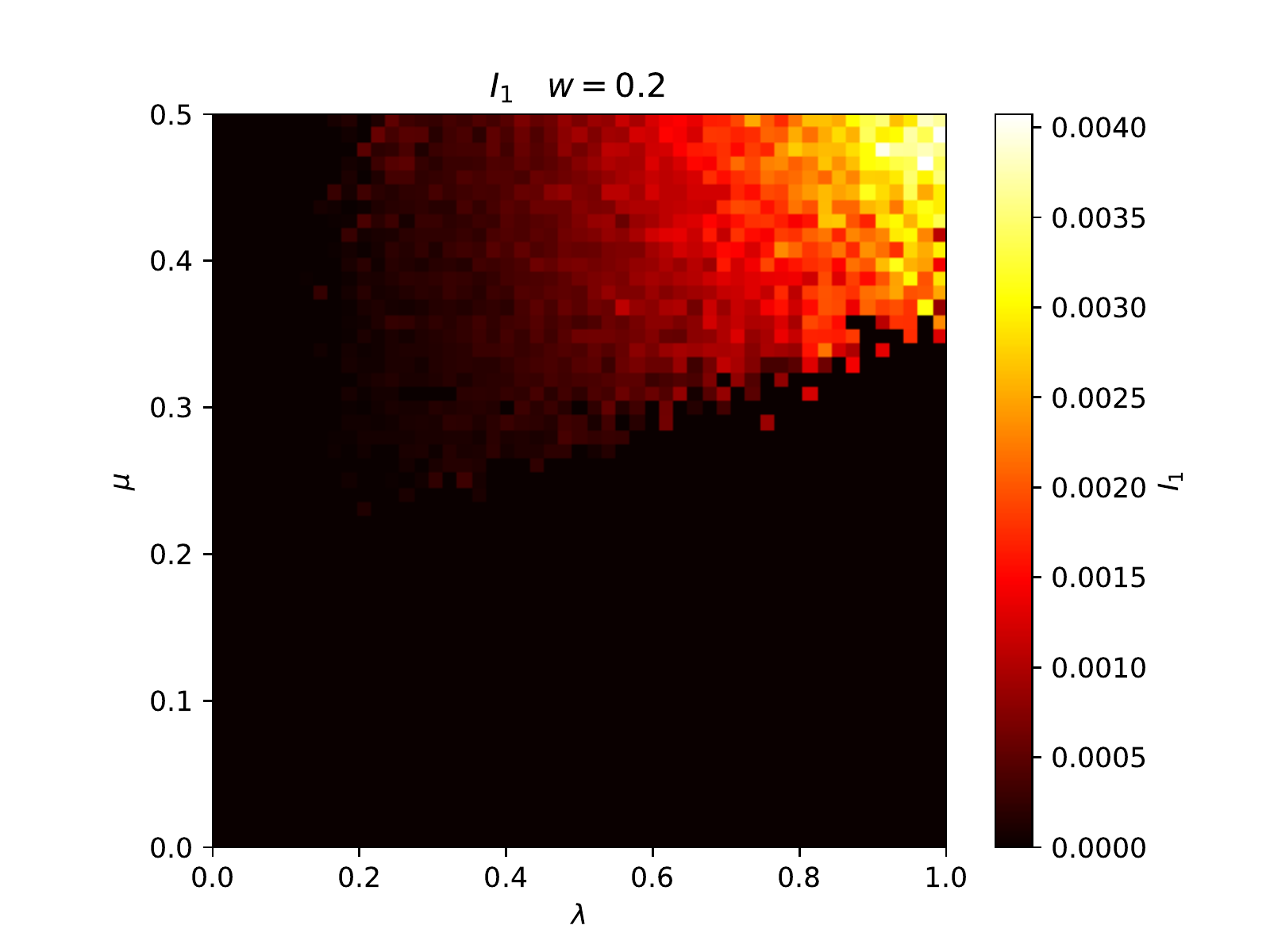}
    \includegraphics[width=0.49\textwidth]{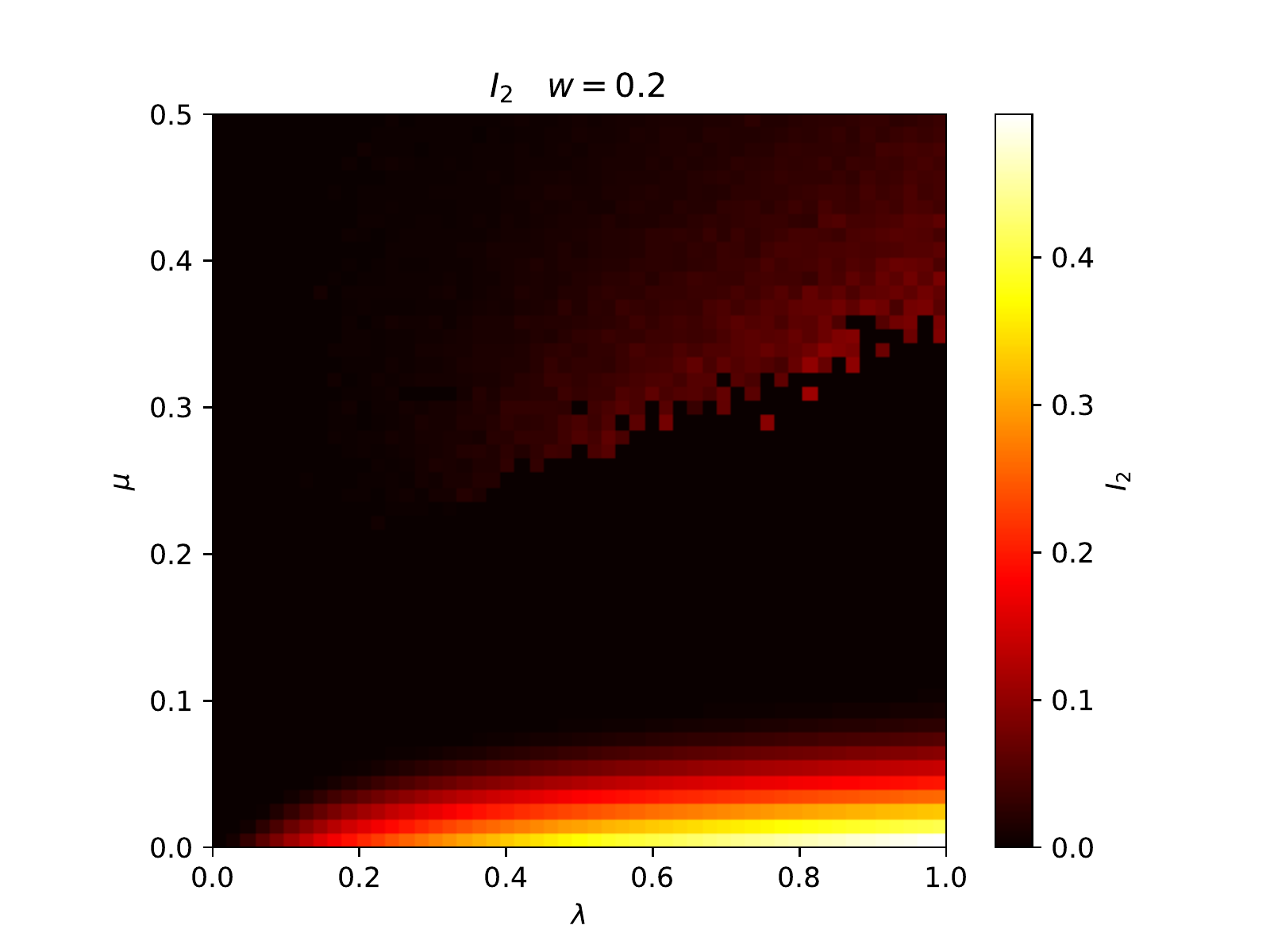}
    \caption{Steady-state for the spreading measure $I_i$ for community 1, above, and 2, bellow. The colors indicate the average of non zero steady-state outcomes of all 200 samples. Results for $w=0.2$.}
    \label{fig:new_results2}
\end{figure}

In this section, we present our results come from Monte Carlo simulations of networks with $N=10^4$ nodes and k=20.
 In all simulations, we set $\alpha=0.1$ and $\phi=0.01$, without loss of generality.
In Figs.~\ref{fig:steadystate-1}-\ref{fig:steadystate-4}, we show the steady-state density of infected agents in the community $u$, $I_u$. We also depict the behavior of the stationary opinion in the community $u$, $m_u$. In turn, $I_{tot}$ and $m_{tot}$ refer to the global proportion of infected individuals and global mean opinion, respectively.

The results in Fig.~\ref{fig:steadystate-1} show that in the community $2$ --- the seed community --- there is a transition from the absorbing phase (extinction of the epidemic) to the epidemic survival phase.
In the community $1$, there is no survival of the chain of infections in the long term.
In this setting with $\mu=0.1$ -- which can be understood as yielding a weak modular structure because of the small value of the parameter -- the seed community remains with the negative opinion about vaccination, which weakens the vaccination campaign and thus facilitates the local permanence of the disease. Similarly, there is a persistence of the initial opinion in the community $1$, which in this case is pro-vaccine and therefore favors the vaccine uptake that makes the epidemic spreading unsustainable. This means that a low number of intercommunity ties hinders the change in the community stance over vaccination; that creates a strong distinction in the epidemic spread between both communities with community $1$ being unfavorable to epidemic spreading since $m_1>0$, and community $2$ being favorable since $m_2<0$.

In Fig.~\ref{fig:steadystate-2}, it is notable that an intermediate community strength leads to the elimination of the epidemic transmission in both communities even when there is a dominance of the negative opinion about vaccination in the community $2$. The epidemic contagion spreading is halted in the community $2$, even though the agents have a negative opinion about the vaccination, due to the intermediate number of bridges, $\mu=0.2$, to the other community. These bridges are just strong enough to drain the infected agents of the community $2$, but not strong enough to change its average opinion.

In Fig.~\ref{fig:steadystate-3}, with $\mu=0.3$ there is a high number of intercommunity links. This additional connectivity between communities weakens the initial epidemic spreading in the community $2$, but it is sufficient to introduce the possibility of a wide opinion change in the community $1$. The opinion change in the community $1$ facilitates the epidemic spreading in that community. This effect is limited because we can see for high infection probabilities $\lambda > 0.8$ the epidemic spread vanishes. So, we have a counterintuitive effect, because for higher transmissibility the epidemic spread vanishes. The reason behind this is the risk perception, $wI$ in Eq.~\eqref{eq:evol}, which promotes vaccination, so higher transmissibility leads to a bigger outbreak that in turn results in better opinions about vaccination which ends up stopping the epidemic outbreak.

The emergence of an intermediate range of $\mu$ that blocks the local and global epidemic spreading is visible in Fig.~\ref{fig:steadystate-4}. Regarding the opinion dynamics, an initial increase in $\mu$ leads to a decrease in $m_1$ and an increase in $m_2$, that is the collective
opinions tend to be less extremist for an initial rise in the amount of inter-communities routes.
Then a further increase in $\mu$ promotes a sudden rise in $m_1$ and $m_2$ which means a speed up in the switch of opinions in the community $2$.
A further rise in $\mu$ leads to a bistable behavior in both communities.

This intermediary range of inter-community connectivity that promotes a minimal epidemic spreading seems to also come from a perceived increment in the probability of an infected individual having a vaccinated neighbor. The increment of the bridges between communities the initially infected agents have a bigger probability of having a neighbor that was vaccinated because initially most of the infected people are in the community $2$ and most agents with a positive opinion about vaccination are in the community $1$. This effect does not persist for higher values of $\mu$ because then both communities tend to adopt the same average opinion about vaccination and this opinion can, in some cases, be negative. A negative global opinion about vaccines does not guarantee that the epidemic spread will persist, as can be seen in some cases for $\mu \approx 0.23$ where all samples had no infected individual but some of them had negative opinions about the vaccination. This can occur due to the fact that the number of infected agents can become zero before the negative global consensus about vaccines is reached.

While in Figs.~\ref{fig:steadystate-1}-\ref{fig:steadystate-2} there is a single stable steady-state (either extinction or persistence), Fig.~\ref{fig:steadystate-3} displays bistable solutions depending on the randomness 'embedded' in the dynamics.
Moreover, the results in Fig.~\ref{fig:steadystate-1} suggest the absorbing-active epidemic transition is continuous for strong communities (such as $\mu=0.1$) whereas the results shown in Fig.~\ref{fig:steadystate-3} signalize this extinction-persistence epidemic transition is discontinuous for weak communities (such as $\mu=0.3$). Therefore, the structural factors present in the modular networks can induce the emergence
of bistability in the epidemic-vaccination-opinion dynamics as well as a change in the nature of the  absorbing-active transitions.

An overall look into Figs.~\ref{fig:steadystate-1}-\ref{fig:steadystate-4} reveals a sudden transition can emerge from structural factors (increasing $\mu$) or epidemiological factors (increasing $\lambda$).
The transitions from the Disease-Free phase to the active phase and vice-versa (epidemic resurgence) highlight the
nonmonotonic behavior of the full dynamics with the transmissibility $\lambda$.

Comparing with other works, we see that while in \cite{2014nematzadehFFA} there is an optimal modularity for enhancing information spreading, here  there is an optimal modularity for hindering epidemic spreading.

In  \cref{fig:new_results1,fig:new_results2} we can see a wide range of results for two different settings of risk perception, i.e. $w = 0.1$ and $w=0.2$. These results are similar but they show how increasing the risk perception reduces the range of parameters that present an endemic state. Other than that, we can also see that in community 1 the endemic state is more prevalent for higher values of modularity $\mu$, this is to be expected since initially only community 2 has infected agents. In community 2 the increment in modularity initially reduces the fraction of infected agents, but at a certain point when the endemic state appears in community 1 it surges back in community 2. This further reinforces that optimal modularity reduces the epidemic spreading.

\FloatBarrier
\section{\label{sec:final}Final Remarks}

In previous work, namely Ref.~\cite{salathe2008effect}, it was shown with a binary opinion dynamics that the spread of opinions against vaccination is one of the potential responsible for the large outbreaks of vaccine-preventable diseases in many high-income countries. In this work, we have gone farther afield to show the emergence of a networked  SIRSV model that the spectrum of scenarios arising from the competition of pro- vs anti-vaccine views during an epidemic spreading is highly complex.

The several outcomes shown in Figs.\ref{fig:steadystate-1}-\ref{fig:new_results2} point out that our model produces a diverse
phenomenology where the social and biological scenarios exhibit a nonmonotonic dependence with spreading rate $\lambda$.  From the
perspective of the  dynamical systems, our results provide a new mechanism for bistability in a
biological-social setting. From a practical point of view, our work offers new perspectives for the development of novel strategies for halting epidemic spreading based on tuning the modularity to an optimal degree. 

Some pro-vaccine strategies can have as side effect the segregation between individuals with conflicting views about the vaccines and clustering of similars. In \cite{kadelka2020effect} the authors found that in scenarios with effective vaccines, the impact of clustering and correlation of belief systems become stronger.
Alternatively, the authors in Ref.~\cite{bizzarri2021epidemic} shown that segregation of anti-vaxxers can potentially extend the duration of an epidemic spreading, whereas in Ref.~\cite{saad2020immune} it  was found that an increase in the contact between vaccine refusers and the rest of the society can lead to a scenario where vaccination alone may not be able to prevent an outbreak. Here we show that too much or too low segregation of anti-vaxxers favors the chain of contagion, but  an intermediate level of segregation  disfavor the epidemic spreading. Therefore, our results indicate that vaccination campaigns should avoid strategies that have as a side effect too much informational segregation of anti-vaccine groups so that reliable pro-vaxx information can reach those groups whilst enforcing a minimum degree of physical distancing as it occurs in countries where childhood vaccination is required at some degree, namely school entry \cite{ourworldindata}.


Our work produces a thought-provoking analogy. In a small-world architecture, there is an intermediate number of long-range bridges that lead the full network to have unusual properties such as high clustering and low path lengths. Here, a structure with an intermediate number of inter-community ties leads the dynamics in the full network to produce an interesting outcome, namely the suppression of the epidemics. Thus, it would be interesting to consider further sophisticated network architectures, like multiplex networks.

Despite the rich phenomenology we observed in our model, some limitations can be discussed which can be targeted in future work. The structure social contacts' structure of modular networks, presenting communities, is relevant to study several dynamical processes \cite{2014nematzadehFFA}-\cite{valdez2020epidemic}. However, it could be more realistic to consider two distinct layers, one for the spreading of each dynamics (epidemic and opinion ones), but with each dynamics influencing the other. Such multiplex network structure can model better the coupled opinion-epidemic dynamics. Other rules for the opinion dynamics, distinct of the kinetic exchanges, could also be considered.

 Besides, it will be  worthwhile to consider the interplay between several sources of heterogeneity in agent’s bias, namely plurality and polarization~\cite{oestereich2020hysteresis}.

\section*{Acknowledgments}

The  authors  acknowledge  financial  support  from  the  Brazilian  funding  agencies  Conselho Nacional de Desenvolvimento Cient\'ifico e Tecnol\'ogico (CNPq), Coordena\c{c}\~ao de Aperfei\c{c}oamento de Pessoal de N\'ivel Superior (CAPES) and Funda\c{c}\~ao Carlos Chagas Filho de Amparo \`a Pesquisa do Estado do Rio de Janeiro (FAPERJ).

\FloatBarrier
\bibliography{apssamp}

\providecommand{\noopsort}[1]{}\providecommand{\singleletter}[1]{#1}%
\begin{thebibliography}{60}%
\makeatletter
\providecommand \@ifxundefined [1]{%
 \@ifx{#1\undefined}
}%
\providecommand \@ifnum [1]{%
 \ifnum #1\expandafter \@firstoftwo
 \else \expandafter \@secondoftwo
 \fi
}%
\providecommand \@ifx [1]{%
 \ifx #1\expandafter \@firstoftwo
 \else \expandafter \@secondoftwo
 \fi
}%
\providecommand \natexlab [1]{#1}%
\providecommand \enquote  [1]{``#1''}%
\providecommand \bibnamefont  [1]{#1}%
\providecommand \bibfnamefont [1]{#1}%
\providecommand \citenamefont [1]{#1}%
\providecommand \href@noop [0]{\@secondoftwo}%
\providecommand \href [0]{\begingroup \@sanitize@url \@href}%
\providecommand \@href[1]{\@@startlink{#1}\@@href}%
\providecommand \@@href[1]{\endgroup#1\@@endlink}%
\providecommand \@sanitize@url [0]{\catcode `\\12\catcode `\$12\catcode
  `\&12\catcode `\#12\catcode `\^12\catcode `\_12\catcode `\%12\relax}%
\providecommand \@@startlink[1]{}%
\providecommand \@@endlink[0]{}%
\providecommand \url  [0]{\begingroup\@sanitize@url \@url }%
\providecommand \@url [1]{\endgroup\@href {#1}{\urlprefix }}%
\providecommand \urlprefix  [0]{URL }%
\providecommand \Eprint [0]{\href }%
\providecommand \doibase [0]{https://doi.org/}%
\providecommand \selectlanguage [0]{\@gobble}%
\providecommand \bibinfo  [0]{\@secondoftwo}%
\providecommand \bibfield  [0]{\@secondoftwo}%
\providecommand \translation [1]{[#1]}%
\providecommand \BibitemOpen [0]{}%
\providecommand \bibitemStop [0]{}%
\providecommand \bibitemNoStop [0]{.\EOS\space}%
\providecommand \EOS [0]{\spacefactor3000\relax}%
\providecommand \BibitemShut  [1]{\csname bibitem#1\endcsname}%
\let\auto@bib@innerbib\@empty
\bibitem [{\citenamefont {Parisi}(1999)}]{parisi1999}%
  \BibitemOpen
  \bibfield  {author} {\bibinfo {author} {\bibfnamefont {G.}~\bibnamefont
  {Parisi}},\ }\bibfield  {title} {\bibinfo {title} {Complex systems: a
  physicist's viewpoint},\ }\href@noop {} {\bibfield  {journal} {\bibinfo
  {journal} {Physica A}\ }\textbf {\bibinfo {volume} {263}},\ \bibinfo {pages}
  {557} (\bibinfo {year} {1999})}\BibitemShut {NoStop}%
\bibitem [{\citenamefont {Thurner}\ \emph {et~al.}(2018)\citenamefont
  {Thurner}, \citenamefont {Hanel},\ and\ \citenamefont
  {Klimek}}]{thurner2018}%
  \BibitemOpen
  \bibfield  {author} {\bibinfo {author} {\bibfnamefont {S.}~\bibnamefont
  {Thurner}}, \bibinfo {author} {\bibfnamefont {R.}~\bibnamefont {Hanel}},\
  and\ \bibinfo {author} {\bibfnamefont {P.}~\bibnamefont {Klimek}},\
  }\href@noop {} {\emph {\bibinfo {title} {Introduction to the Theory of
  Complex Systems}}}\ (\bibinfo  {publisher} {Oxford, University Press},\
  \bibinfo {address} {Oxford},\ \bibinfo {year} {2018})\BibitemShut {NoStop}%
\bibitem [{\citenamefont {Stauffer}\ \emph {et~al.}(2006)\citenamefont
  {Stauffer}, \citenamefont {De~Oliveira}, \citenamefont {De~Oliveira},\ and\
  \citenamefont {de~S{\'a}~Martins}}]{stauffer2006biology}%
  \BibitemOpen
  \bibfield  {author} {\bibinfo {author} {\bibfnamefont {D.}~\bibnamefont
  {Stauffer}}, \bibinfo {author} {\bibfnamefont {S.~M.~M.}\ \bibnamefont
  {De~Oliveira}}, \bibinfo {author} {\bibfnamefont {P.~M.~C.}\ \bibnamefont
  {De~Oliveira}},\ and\ \bibinfo {author} {\bibfnamefont {J.~S.}\ \bibnamefont
  {de~S{\'a}~Martins}},\ }\href@noop {} {\emph {\bibinfo {title} {Biology,
  sociology, geology by computational physicists}}}\ (\bibinfo  {publisher}
  {Elsevier},\ \bibinfo {year} {2006})\BibitemShut {NoStop}%
\bibitem [{\citenamefont {Castellano}\ \emph {et~al.}(2009)\citenamefont
  {Castellano}, \citenamefont {Fortunato},\ and\ \citenamefont
  {Loreto}}]{castellano2009statistical}%
  \BibitemOpen
  \bibfield  {author} {\bibinfo {author} {\bibfnamefont {C.}~\bibnamefont
  {Castellano}}, \bibinfo {author} {\bibfnamefont {S.}~\bibnamefont
  {Fortunato}},\ and\ \bibinfo {author} {\bibfnamefont {V.}~\bibnamefont
  {Loreto}},\ }\bibfield  {title} {\bibinfo {title} {Statistical physics of
  social dynamics},\ }\href@noop {} {\bibfield  {journal} {\bibinfo  {journal}
  {Reviews of modern physics}\ }\textbf {\bibinfo {volume} {81}},\ \bibinfo
  {pages} {591} (\bibinfo {year} {2009})}\BibitemShut {NoStop}%
\bibitem [{\citenamefont {de~Oliveira}\ and\ \citenamefont
  {Stauffer}(2013)}]{de2013evolution}%
  \BibitemOpen
  \bibfield  {author} {\bibinfo {author} {\bibfnamefont {P.~M.~C.}\
  \bibnamefont {de~Oliveira}}\ and\ \bibinfo {author} {\bibfnamefont
  {D.}~\bibnamefont {Stauffer}},\ }\href@noop {} {\emph {\bibinfo {title}
  {Evolution, Money, War, and Computers: Non-Traditional Applications of
  Computational Statistical Physics}}},\ Vol.~\bibinfo {volume} {34}\ (\bibinfo
   {publisher} {Springer Science \& Business Media},\ \bibinfo {year}
  {2013})\BibitemShut {NoStop}%
\bibitem [{\citenamefont {Galam}(2008)}]{galam2008sociophysics}%
  \BibitemOpen
  \bibfield  {author} {\bibinfo {author} {\bibfnamefont {S.}~\bibnamefont
  {Galam}},\ }\bibfield  {title} {\bibinfo {title} {Sociophysics: A review of
  galam models},\ }\href@noop {} {\bibfield  {journal} {\bibinfo  {journal}
  {International Journal of Modern Physics C}\ }\textbf {\bibinfo {volume}
  {19}},\ \bibinfo {pages} {409} (\bibinfo {year} {2008})}\BibitemShut
  {NoStop}%
\bibitem [{\citenamefont {Sen}\ and\ \citenamefont
  {Chakrabarti}(2014)}]{sen2014sociophysics}%
  \BibitemOpen
  \bibfield  {author} {\bibinfo {author} {\bibfnamefont {P.}~\bibnamefont
  {Sen}}\ and\ \bibinfo {author} {\bibfnamefont {B.~K.}\ \bibnamefont
  {Chakrabarti}},\ }\href@noop {} {\emph {\bibinfo {title} {Sociophysics: an
  introduction}}}\ (\bibinfo  {publisher} {Oxford University Press},\ \bibinfo
  {year} {2014})\BibitemShut {NoStop}%
\bibitem [{\citenamefont {Pastor-Satorras}\ \emph {et~al.}(2015)\citenamefont
  {Pastor-Satorras}, \citenamefont {Van~Mieghem},\ and\ \citenamefont
  {Vespignani}}]{pastor2015}%
  \BibitemOpen
  \bibfield  {author} {\bibinfo {author} {\bibfnamefont {R.}~\bibnamefont
  {Pastor-Satorras}}, \bibinfo {author} {\bibfnamefont {P.}~\bibnamefont
  {Van~Mieghem}},\ and\ \bibinfo {author} {\bibfnamefont {A.}~\bibnamefont
  {Vespignani}},\ }\bibfield  {title} {\bibinfo {title} {Epidemic processes in
  complex networks},\ }\href@noop {} {\bibfield  {journal} {\bibinfo  {journal}
  {Rev. Mod. Phys.}\ }\textbf {\bibinfo {volume} {87}},\ \bibinfo {pages} {925}
  (\bibinfo {year} {2015})}\BibitemShut {NoStop}%
\bibitem [{\citenamefont {de~Arruda}\ \emph {et~al.}(2018)\citenamefont
  {de~Arruda}, \citenamefont {Rodrigues},\ and\ \citenamefont
  {Moreno}}]{arruda2018}%
  \BibitemOpen
  \bibfield  {author} {\bibinfo {author} {\bibfnamefont {G.~F.}\ \bibnamefont
  {de~Arruda}}, \bibinfo {author} {\bibfnamefont {F.~A.}\ \bibnamefont
  {Rodrigues}},\ and\ \bibinfo {author} {\bibfnamefont {Y.}~\bibnamefont
  {Moreno}},\ }\bibfield  {title} {\bibinfo {title} {Fundamentals of spreading
  processes in single and multilayer complex networks},\ }\href@noop {}
  {\bibfield  {journal} {\bibinfo  {journal} {Phys. Rep.}\ }\textbf {\bibinfo
  {volume} {756}},\ \bibinfo {pages} {1} (\bibinfo {year} {2018})}\BibitemShut
  {NoStop}%
\bibitem [{\citenamefont {Wang}\ \emph {et~al.}(2017)\citenamefont {Wang},
  \citenamefont {Tang}, \citenamefont {Braunstein},\ and\ \citenamefont
  {Stanley}}]{wang2017}%
  \BibitemOpen
  \bibfield  {author} {\bibinfo {author} {\bibfnamefont {W.}~\bibnamefont
  {Wang}}, \bibinfo {author} {\bibfnamefont {M.}~\bibnamefont {Tang}}, \bibinfo
  {author} {\bibfnamefont {L.~A.}\ \bibnamefont {Braunstein}},\ and\ \bibinfo
  {author} {\bibfnamefont {E.~H.}\ \bibnamefont {Stanley}},\ }\bibfield
  {title} {\bibinfo {title} {Unification of theoretical approaches for epidemic
  spreading on complex networks},\ }\href@noop {} {\bibfield  {journal}
  {\bibinfo  {journal} {Rep. Prog. Phys.}\ }\textbf {\bibinfo {volume} {80}},\
  \bibinfo {pages} {036603} (\bibinfo {year} {2017})}\BibitemShut {NoStop}%
\bibitem [{\citenamefont {Wang}\ \emph {et~al.}(2019)\citenamefont {Wang},
  \citenamefont {Liu}, \citenamefont {Liang}, \citenamefont {Hu},\ and\
  \citenamefont {Zhou}}]{wang2019coevolution}%
  \BibitemOpen
  \bibfield  {author} {\bibinfo {author} {\bibfnamefont {W.}~\bibnamefont
  {Wang}}, \bibinfo {author} {\bibfnamefont {Q.-H.}\ \bibnamefont {Liu}},
  \bibinfo {author} {\bibfnamefont {J.}~\bibnamefont {Liang}}, \bibinfo
  {author} {\bibfnamefont {Y.}~\bibnamefont {Hu}},\ and\ \bibinfo {author}
  {\bibfnamefont {T.}~\bibnamefont {Zhou}},\ }\bibfield  {title} {\bibinfo
  {title} {Coevolution spreading in complex networks},\ }\bibfield  {journal}
  {\bibinfo  {journal} {Physics Reports}\ }\href
  {https://doi.org/10.1016/j.physrep.2019.07.001}
  {10.1016/j.physrep.2019.07.001} (\bibinfo {year} {2019})\BibitemShut
  {NoStop}%
\bibitem [{\citenamefont {Gasparini}\ \emph {et~al.}(2015)\citenamefont
  {Gasparini}, \citenamefont {Panatto}, \citenamefont {Lai},\ and\
  \citenamefont {Amicizia}}]{gasparini2015}%
  \BibitemOpen
  \bibfield  {author} {\bibinfo {author} {\bibfnamefont {R.}~\bibnamefont
  {Gasparini}}, \bibinfo {author} {\bibfnamefont {D.}~\bibnamefont {Panatto}},
  \bibinfo {author} {\bibfnamefont {P.~L.}\ \bibnamefont {Lai}},\ and\ \bibinfo
  {author} {\bibfnamefont {D.}~\bibnamefont {Amicizia}},\ }\bibfield  {title}
  {\bibinfo {title} {The ``urban myth'' of the association between neurological
  disorders and vaccinations},\ }\href@noop {} {\bibfield  {journal} {\bibinfo
  {journal} {J. Prev. Med. Hyg.}\ }\textbf {\bibinfo {volume} {56}},\ \bibinfo
  {pages} {e1} (\bibinfo {year} {2015})}\BibitemShut {NoStop}%
\bibitem [{\citenamefont {Feemster}\ and\ \citenamefont
  {Szipszky}(2020)}]{feemster2020}%
  \BibitemOpen
  \bibfield  {author} {\bibinfo {author} {\bibfnamefont {K.~A.}\ \bibnamefont
  {Feemster}}\ and\ \bibinfo {author} {\bibfnamefont {C.}~\bibnamefont
  {Szipszky}},\ }\bibfield  {title} {\bibinfo {title} {Resurgence of measles in
  the united states: how did we get here?},\ }\href@noop {} {\bibfield
  {journal} {\bibinfo  {journal} {Curr. Opin. Pediatr.}\ }\textbf {\bibinfo
  {volume} {32}},\ \bibinfo {pages} {139} (\bibinfo {year} {2020})}\BibitemShut
  {NoStop}%
\bibitem [{\citenamefont {Nematzadeh}\ \emph {et~al.}(2014)\citenamefont
  {Nematzadeh}, \citenamefont {Ferrara}, \citenamefont {Flammini},\ and\
  \citenamefont {Ahn}}]{2014nematzadehFFA}%
  \BibitemOpen
  \bibfield  {author} {\bibinfo {author} {\bibfnamefont {A.}~\bibnamefont
  {Nematzadeh}}, \bibinfo {author} {\bibfnamefont {E.}~\bibnamefont {Ferrara}},
  \bibinfo {author} {\bibfnamefont {A.}~\bibnamefont {Flammini}},\ and\
  \bibinfo {author} {\bibfnamefont {Y.-Y.}\ \bibnamefont {Ahn}},\ }\bibfield
  {title} {\bibinfo {title} {Optimal network modularity for information
  diffusion},\ }\href {https://doi.org/10.1103/PhysRevLett.113.088701}
  {\bibfield  {journal} {\bibinfo  {journal} {Phys. Rev. Lett.}\ }\textbf
  {\bibinfo {volume} {113}},\ \bibinfo {pages} {088701} (\bibinfo {year}
  {2014})}\BibitemShut {NoStop}%
\bibitem [{\citenamefont {Su}\ \emph {et~al.}(2018)\citenamefont {Su},
  \citenamefont {Wang}, \citenamefont {Li}, \citenamefont {Stanley},\ and\
  \citenamefont {Braunstein}}]{su2018optimal}%
  \BibitemOpen
  \bibfield  {author} {\bibinfo {author} {\bibfnamefont {Z.}~\bibnamefont
  {Su}}, \bibinfo {author} {\bibfnamefont {W.}~\bibnamefont {Wang}}, \bibinfo
  {author} {\bibfnamefont {L.}~\bibnamefont {Li}}, \bibinfo {author}
  {\bibfnamefont {H.~E.}\ \bibnamefont {Stanley}},\ and\ \bibinfo {author}
  {\bibfnamefont {L.~A.}\ \bibnamefont {Braunstein}},\ }\bibfield  {title}
  {\bibinfo {title} {Optimal community structure for social contagions},\
  }\href@noop {} {\bibfield  {journal} {\bibinfo  {journal} {New Journal of
  Physics}\ }\textbf {\bibinfo {volume} {20}},\ \bibinfo {pages} {053053}
  (\bibinfo {year} {2018})}\BibitemShut {NoStop}%
\bibitem [{\citenamefont {Wu}\ \emph {et~al.}(2016{\natexlab{a}})\citenamefont
  {Wu}, \citenamefont {Du}, \citenamefont {Zheng},\ and\ \citenamefont
  {Liu}}]{wu2016optimal}%
  \BibitemOpen
  \bibfield  {author} {\bibinfo {author} {\bibfnamefont {J.}~\bibnamefont
  {Wu}}, \bibinfo {author} {\bibfnamefont {R.}~\bibnamefont {Du}}, \bibinfo
  {author} {\bibfnamefont {Y.}~\bibnamefont {Zheng}},\ and\ \bibinfo {author}
  {\bibfnamefont {D.}~\bibnamefont {Liu}},\ }\bibfield  {title} {\bibinfo
  {title} {Optimal multi-community network modularity for information
  diffusion},\ }\href@noop {} {\bibfield  {journal} {\bibinfo  {journal}
  {International Journal of Modern Physics C}\ }\textbf {\bibinfo {volume}
  {27}},\ \bibinfo {pages} {1650092} (\bibinfo {year}
  {2016}{\natexlab{a}})}\BibitemShut {NoStop}%
\bibitem [{\citenamefont {Romano}\ \emph {et~al.}(2018)\citenamefont {Romano},
  \citenamefont {Shen}, \citenamefont {Pansanel}, \citenamefont {MacIntosh},\
  and\ \citenamefont {Sueur}}]{romano2018social}%
  \BibitemOpen
  \bibfield  {author} {\bibinfo {author} {\bibfnamefont {V.}~\bibnamefont
  {Romano}}, \bibinfo {author} {\bibfnamefont {M.}~\bibnamefont {Shen}},
  \bibinfo {author} {\bibfnamefont {J.}~\bibnamefont {Pansanel}}, \bibinfo
  {author} {\bibfnamefont {A.~J.}\ \bibnamefont {MacIntosh}},\ and\ \bibinfo
  {author} {\bibfnamefont {C.}~\bibnamefont {Sueur}},\ }\bibfield  {title}
  {\bibinfo {title} {Social transmission in networks: global efficiency peaks
  with intermediate levels of modularity},\ }\href@noop {} {\bibfield
  {journal} {\bibinfo  {journal} {Behavioral ecology and sociobiology}\
  }\textbf {\bibinfo {volume} {72}},\ \bibinfo {pages} {154} (\bibinfo {year}
  {2018})}\BibitemShut {NoStop}%
\bibitem [{\citenamefont {Rodriguez}\ \emph {et~al.}(2019)\citenamefont
  {Rodriguez}, \citenamefont {Izquierdo},\ and\ \citenamefont
  {Ahn}}]{rodriguez2019optimal}%
  \BibitemOpen
  \bibfield  {author} {\bibinfo {author} {\bibfnamefont {N.}~\bibnamefont
  {Rodriguez}}, \bibinfo {author} {\bibfnamefont {E.}~\bibnamefont
  {Izquierdo}},\ and\ \bibinfo {author} {\bibfnamefont {Y.-Y.}\ \bibnamefont
  {Ahn}},\ }\bibfield  {title} {\bibinfo {title} {Optimal modularity and memory
  capacity of neural reservoirs},\ }\href@noop {} {\bibfield  {journal}
  {\bibinfo  {journal} {Network Neuroscience}\ }\textbf {\bibinfo {volume}
  {3}},\ \bibinfo {pages} {551} (\bibinfo {year} {2019})}\BibitemShut {NoStop}%
\bibitem [{\citenamefont {Cui}\ \emph {et~al.}(2018)\citenamefont {Cui},
  \citenamefont {Wang}, \citenamefont {Cai}, \citenamefont {Zhou},
  \citenamefont {Lai} \emph {et~al.}}]{cui2018close}%
  \BibitemOpen
  \bibfield  {author} {\bibinfo {author} {\bibfnamefont {P.-B.}\ \bibnamefont
  {Cui}}, \bibinfo {author} {\bibfnamefont {W.}~\bibnamefont {Wang}}, \bibinfo
  {author} {\bibfnamefont {S.-M.}\ \bibnamefont {Cai}}, \bibinfo {author}
  {\bibfnamefont {T.}~\bibnamefont {Zhou}}, \bibinfo {author} {\bibfnamefont
  {Y.-C.}\ \bibnamefont {Lai}}, \emph {et~al.},\ }\bibfield  {title} {\bibinfo
  {title} {Close and ordinary social contacts: How important are they in
  promoting large-scale contagion?},\ }\href@noop {} {\bibfield  {journal}
  {\bibinfo  {journal} {Physical Review E}\ }\textbf {\bibinfo {volume} {98}},\
  \bibinfo {pages} {052311} (\bibinfo {year} {2018})}\BibitemShut {NoStop}%
\bibitem [{\citenamefont {Curato}\ and\ \citenamefont
  {Lillo}(2016)}]{curato2016optimal}%
  \BibitemOpen
  \bibfield  {author} {\bibinfo {author} {\bibfnamefont {G.}~\bibnamefont
  {Curato}}\ and\ \bibinfo {author} {\bibfnamefont {F.}~\bibnamefont {Lillo}},\
  }\bibfield  {title} {\bibinfo {title} {Optimal information diffusion in
  stochastic block models},\ }\href@noop {} {\bibfield  {journal} {\bibinfo
  {journal} {Physical Review E}\ }\textbf {\bibinfo {volume} {94}},\ \bibinfo
  {pages} {032310} (\bibinfo {year} {2016})}\BibitemShut {NoStop}%
\bibitem [{\citenamefont {Nematzadeh}\ \emph {et~al.}(2018)\citenamefont
  {Nematzadeh}, \citenamefont {Rodriguez}, \citenamefont {Flammini},\ and\
  \citenamefont {Ahn}}]{nematzadeh2018optimal}%
  \BibitemOpen
  \bibfield  {author} {\bibinfo {author} {\bibfnamefont {A.}~\bibnamefont
  {Nematzadeh}}, \bibinfo {author} {\bibfnamefont {N.}~\bibnamefont
  {Rodriguez}}, \bibinfo {author} {\bibfnamefont {A.}~\bibnamefont
  {Flammini}},\ and\ \bibinfo {author} {\bibfnamefont {Y.-Y.}\ \bibnamefont
  {Ahn}},\ }\bibfield  {title} {\bibinfo {title} {Optimal modularity in complex
  contagion},\ }in\ \href@noop {} {\emph {\bibinfo {booktitle} {Complex
  spreading phenomena in social systems}}}\ (\bibinfo  {publisher} {Springer},\
  \bibinfo {year} {2018})\ pp.\ \bibinfo {pages} {97--107}\BibitemShut
  {NoStop}%
\bibitem [{\citenamefont {Peng}\ \emph {et~al.}(2020)\citenamefont {Peng},
  \citenamefont {Nematzadeh}, \citenamefont {Romero},\ and\ \citenamefont
  {Ferrara}}]{PhysRevE.102.052316}%
  \BibitemOpen
  \bibfield  {author} {\bibinfo {author} {\bibfnamefont {H.}~\bibnamefont
  {Peng}}, \bibinfo {author} {\bibfnamefont {A.}~\bibnamefont {Nematzadeh}},
  \bibinfo {author} {\bibfnamefont {D.~M.}\ \bibnamefont {Romero}},\ and\
  \bibinfo {author} {\bibfnamefont {E.}~\bibnamefont {Ferrara}},\ }\bibfield
  {title} {\bibinfo {title} {Network modularity controls the speed of
  information diffusion},\ }\href {https://doi.org/10.1103/PhysRevE.102.052316}
  {\bibfield  {journal} {\bibinfo  {journal} {Phys. Rev. E}\ }\textbf {\bibinfo
  {volume} {102}},\ \bibinfo {pages} {052316} (\bibinfo {year}
  {2020})}\BibitemShut {NoStop}%
\bibitem [{\citenamefont {Valdez}\ \emph {et~al.}(2020)\citenamefont {Valdez},
  \citenamefont {Braunstein},\ and\ \citenamefont
  {Havlin}}]{valdez2020epidemic}%
  \BibitemOpen
  \bibfield  {author} {\bibinfo {author} {\bibfnamefont {L.~D.}\ \bibnamefont
  {Valdez}}, \bibinfo {author} {\bibfnamefont {L.~A.}\ \bibnamefont
  {Braunstein}},\ and\ \bibinfo {author} {\bibfnamefont {S.}~\bibnamefont
  {Havlin}},\ }\bibfield  {title} {\bibinfo {title} {Epidemic spreading on
  modular networks: The fear to declare a pandemic},\ }\href@noop {} {\bibfield
   {journal} {\bibinfo  {journal} {Physical Review E}\ }\textbf {\bibinfo
  {volume} {101}},\ \bibinfo {pages} {032309} (\bibinfo {year}
  {2020})}\BibitemShut {NoStop}%
\bibitem [{\citenamefont {Fu}\ \emph {et~al.}(2011)\citenamefont {Fu},
  \citenamefont {Rosenbloom}, \citenamefont {Wang},\ and\ \citenamefont
  {Nowak}}]{Nowak2011}%
  \BibitemOpen
  \bibfield  {author} {\bibinfo {author} {\bibfnamefont {F.}~\bibnamefont
  {Fu}}, \bibinfo {author} {\bibfnamefont {D.~I.}\ \bibnamefont {Rosenbloom}},
  \bibinfo {author} {\bibfnamefont {L.}~\bibnamefont {Wang}},\ and\ \bibinfo
  {author} {\bibfnamefont {M.~A.}\ \bibnamefont {Nowak}},\ }\bibfield  {title}
  {\bibinfo {title} {Imitation dynamics of vaccination behaviour on social
  networks},\ }\href {https://doi.org/10.1098/rspb.2010.1107} {\bibfield
  {journal} {\bibinfo  {journal} {Proceedings of the Royal Society B:
  Biological Sciences}\ }\textbf {\bibinfo {volume} {278}},\ \bibinfo {pages}
  {42} (\bibinfo {year} {2011})}\BibitemShut {NoStop}%
\bibitem [{\citenamefont {Chen}\ and\ \citenamefont {Fu}(2019)}]{Feng2019}%
  \BibitemOpen
  \bibfield  {author} {\bibinfo {author} {\bibfnamefont {X.}~\bibnamefont
  {Chen}}\ and\ \bibinfo {author} {\bibfnamefont {F.}~\bibnamefont {Fu}},\
  }\bibfield  {title} {\bibinfo {title} {Imperfect vaccine and hysteresis},\
  }\href {https://doi.org/10.1098/rspb.2018.2406} {\bibfield  {journal}
  {\bibinfo  {journal} {Proceedings of the Royal Society B: Biological
  Sciences}\ }\textbf {\bibinfo {volume} {286}},\ \bibinfo {pages} {20182406}
  (\bibinfo {year} {2019})}\BibitemShut {NoStop}%
\bibitem [{\citenamefont {Bi}\ \emph {et~al.}(2019)\citenamefont {Bi},
  \citenamefont {Chen}, \citenamefont {Zhao}, \citenamefont {Ben-Arieh},\ and\
  \citenamefont {Wu}}]{bi2019modeling}%
  \BibitemOpen
  \bibfield  {author} {\bibinfo {author} {\bibfnamefont {K.}~\bibnamefont
  {Bi}}, \bibinfo {author} {\bibfnamefont {Y.}~\bibnamefont {Chen}}, \bibinfo
  {author} {\bibfnamefont {S.}~\bibnamefont {Zhao}}, \bibinfo {author}
  {\bibfnamefont {D.}~\bibnamefont {Ben-Arieh}},\ and\ \bibinfo {author}
  {\bibfnamefont {C.-H.~J.}\ \bibnamefont {Wu}},\ }\bibfield  {title} {\bibinfo
  {title} {Modeling learning and forgetting processes with the corresponding
  impacts on human behaviors in infectious disease epidemics},\ }\href@noop {}
  {\bibfield  {journal} {\bibinfo  {journal} {Computers \& Industrial
  Engineering}\ }\textbf {\bibinfo {volume} {129}},\ \bibinfo {pages} {563}
  (\bibinfo {year} {2019})}\BibitemShut {NoStop}%
\bibitem [{\citenamefont {Zhao}\ \emph {et~al.}(2018)\citenamefont {Zhao},
  \citenamefont {Kuang}, \citenamefont {Wu}, \citenamefont {Bi},\ and\
  \citenamefont {Ben-Arieh}}]{zhao2018risk}%
  \BibitemOpen
  \bibfield  {author} {\bibinfo {author} {\bibfnamefont {S.}~\bibnamefont
  {Zhao}}, \bibinfo {author} {\bibfnamefont {Y.}~\bibnamefont {Kuang}},
  \bibinfo {author} {\bibfnamefont {C.-H.}\ \bibnamefont {Wu}}, \bibinfo
  {author} {\bibfnamefont {K.}~\bibnamefont {Bi}},\ and\ \bibinfo {author}
  {\bibfnamefont {D.}~\bibnamefont {Ben-Arieh}},\ }\bibfield  {title} {\bibinfo
  {title} {Risk perception and human behaviors in epidemics},\ }\href@noop {}
  {\bibfield  {journal} {\bibinfo  {journal} {IISE Transactions on Healthcare
  Systems Engineering}\ }\textbf {\bibinfo {volume} {8}},\ \bibinfo {pages}
  {315} (\bibinfo {year} {2018})}\BibitemShut {NoStop}%
\bibitem [{\citenamefont {Smaldino}\ and\ \citenamefont
  {Jones}(2020)}]{smaldino2020coupled}%
  \BibitemOpen
  \bibfield  {author} {\bibinfo {author} {\bibfnamefont {P.~E.}\ \bibnamefont
  {Smaldino}}\ and\ \bibinfo {author} {\bibfnamefont {J.~H.}\ \bibnamefont
  {Jones}},\ }\bibfield  {title} {\bibinfo {title} {Coupled dynamics of
  behavior and disease contagion among antagonistic groups},\ }\href@noop {}
  {\bibfield  {journal} {\bibinfo  {journal} {bioRxiv}\ } (\bibinfo {year}
  {2020})}\BibitemShut {NoStop}%
\bibitem [{\citenamefont {Galam}(2010)}]{galam2010public}%
  \BibitemOpen
  \bibfield  {author} {\bibinfo {author} {\bibfnamefont {S.}~\bibnamefont
  {Galam}},\ }\bibfield  {title} {\bibinfo {title} {Public debates driven by
  incomplete scientific data: the cases of evolution theory, global warming and
  h1n1 pandemic influenza},\ }\href
  {https://doi.org/10.1016/j.physa.2010.04.039} {\bibfield  {journal} {\bibinfo
   {journal} {Physica A: Statistical Mechanics and its Applications}\ }\textbf
  {\bibinfo {volume} {389}},\ \bibinfo {pages} {3619} (\bibinfo {year}
  {2010})}\BibitemShut {NoStop}%
\bibitem [{\citenamefont {Pires}\ and\ \citenamefont
  {Crokidakis}(2017)}]{pires2017dynamics}%
  \BibitemOpen
  \bibfield  {author} {\bibinfo {author} {\bibfnamefont {M.~A.}\ \bibnamefont
  {Pires}}\ and\ \bibinfo {author} {\bibfnamefont {N.}~\bibnamefont
  {Crokidakis}},\ }\bibfield  {title} {\bibinfo {title} {Dynamics of epidemic
  spreading with vaccination: impact of social pressure and engagement},\
  }\href {https://doi.org/10.1016/j.physa.2016.10.004} {\bibfield  {journal}
  {\bibinfo  {journal} {Physica A}\ }\textbf {\bibinfo {volume} {467}},\
  \bibinfo {pages} {167} (\bibinfo {year} {2017})}\BibitemShut {NoStop}%
\bibitem [{\citenamefont {Pires}\ \emph {et~al.}(2018)\citenamefont {Pires},
  \citenamefont {Oestereich},\ and\ \citenamefont {Crokidakis}}]{2018piresOC}%
  \BibitemOpen
  \bibfield  {author} {\bibinfo {author} {\bibfnamefont {M.~A.}\ \bibnamefont
  {Pires}}, \bibinfo {author} {\bibfnamefont {A.~L.}\ \bibnamefont
  {Oestereich}},\ and\ \bibinfo {author} {\bibfnamefont {N.}~\bibnamefont
  {Crokidakis}},\ }\bibfield  {title} {\bibinfo {title} {Sudden transitions in
  coupled opinion and epidemic dynamics with vaccination},\ }\href
  {https://doi.org/10.1088/1742-5468/aabfc6} {\bibfield  {journal} {\bibinfo
  {journal} {J. Stat. Mech.}\ }\textbf {\bibinfo {volume} {2018}},\ \bibinfo
  {pages} {053407} (\bibinfo {year} {2018})}\BibitemShut {NoStop}%
\bibitem [{\citenamefont {Mehta}\ and\ \citenamefont
  {Rosenberg}(2020)}]{mehta2020modelling}%
  \BibitemOpen
  \bibfield  {author} {\bibinfo {author} {\bibfnamefont {R.~S.}\ \bibnamefont
  {Mehta}}\ and\ \bibinfo {author} {\bibfnamefont {N.~A.}\ \bibnamefont
  {Rosenberg}},\ }\bibfield  {title} {\bibinfo {title} {Modelling anti-vaccine
  sentiment as a cultural pathogen},\ }\href@noop {} {\bibfield  {journal}
  {\bibinfo  {journal} {Evolutionary Human Sciences}\ }\textbf {\bibinfo
  {volume} {2}} (\bibinfo {year} {2020})}\BibitemShut {NoStop}%
\bibitem [{\citenamefont {Burki}(2020)}]{burki2020online}%
  \BibitemOpen
  \bibfield  {author} {\bibinfo {author} {\bibfnamefont {T.}~\bibnamefont
  {Burki}},\ }\bibfield  {title} {\bibinfo {title} {The online anti-vaccine
  movement in the age of covid-19},\ }\href@noop {} {\bibfield  {journal}
  {\bibinfo  {journal} {The Lancet Digital Health}\ }\textbf {\bibinfo {volume}
  {2}},\ \bibinfo {pages} {e504} (\bibinfo {year} {2020})}\BibitemShut
  {NoStop}%
\bibitem [{\citenamefont {Johnson}\ \emph
  {et~al.}(2020{\natexlab{a}})\citenamefont {Johnson}, \citenamefont
  {Vel{\'a}squez}, \citenamefont {Restrepo}, \citenamefont {Leahy},
  \citenamefont {Gabriel}, \citenamefont {El~Oud}, \citenamefont {Zheng},
  \citenamefont {Manrique}, \citenamefont {Wuchty},\ and\ \citenamefont
  {Lupu}}]{johnson2020online}%
  \BibitemOpen
  \bibfield  {author} {\bibinfo {author} {\bibfnamefont {N.~F.}\ \bibnamefont
  {Johnson}}, \bibinfo {author} {\bibfnamefont {N.}~\bibnamefont
  {Vel{\'a}squez}}, \bibinfo {author} {\bibfnamefont {N.~J.}\ \bibnamefont
  {Restrepo}}, \bibinfo {author} {\bibfnamefont {R.}~\bibnamefont {Leahy}},
  \bibinfo {author} {\bibfnamefont {N.}~\bibnamefont {Gabriel}}, \bibinfo
  {author} {\bibfnamefont {S.}~\bibnamefont {El~Oud}}, \bibinfo {author}
  {\bibfnamefont {M.}~\bibnamefont {Zheng}}, \bibinfo {author} {\bibfnamefont
  {P.}~\bibnamefont {Manrique}}, \bibinfo {author} {\bibfnamefont
  {S.}~\bibnamefont {Wuchty}},\ and\ \bibinfo {author} {\bibfnamefont
  {Y.}~\bibnamefont {Lupu}},\ }\bibfield  {title} {\bibinfo {title} {The online
  competition between pro-and anti-vaccination views},\ }\href
  {https://doi.org/10.1038/s41586-020-2281-1} {\bibfield  {journal} {\bibinfo
  {journal} {Nature}\ }\textbf {\bibinfo {volume} {582}},\ \bibinfo {pages}
  {230} (\bibinfo {year} {2020}{\natexlab{a}})}\BibitemShut {NoStop}%
\bibitem [{\citenamefont {Buonomo}(2020)}]{buonomo2020effects}%
  \BibitemOpen
  \bibfield  {author} {\bibinfo {author} {\bibfnamefont {B.}~\bibnamefont
  {Buonomo}},\ }\bibfield  {title} {\bibinfo {title} {Effects of
  information-dependent vaccination behavior on coronavirus outbreak: insights
  from a siri model},\ }\href@noop {} {\bibfield  {journal} {\bibinfo
  {journal} {Ricerche di Matematica}\ }\textbf {\bibinfo {volume} {69}},\
  \bibinfo {pages} {483} (\bibinfo {year} {2020})}\BibitemShut {NoStop}%
\bibitem [{\citenamefont {Boyon}\ and\ \citenamefont
  {Silverstein}(2020)}]{boyon}%
  \BibitemOpen
  \bibfield  {author} {\bibinfo {author} {\bibfnamefont {N.}~\bibnamefont
  {Boyon}}\ and\ \bibinfo {author} {\bibfnamefont {K.}~\bibnamefont
  {Silverstein}},\ }\bibfield  {title} {\bibinfo {title} {Three in four adults
  globally say they would get a vaccine for covid-19},\ }\href
  {https://www.ipsos.com/en-us/news-polls/WEF-covid-vaccine-global} {\bibfield
  {journal} {\bibinfo  {journal} {Ipsos, News \& Pools}\ } (\bibinfo {year}
  {2020})}\BibitemShut {NoStop}%
\bibitem [{\citenamefont {Curiel}\ and\ \citenamefont
  {Ramírez}(2020)}]{curielantivax}%
  \BibitemOpen
  \bibfield  {author} {\bibinfo {author} {\bibfnamefont {R.~P.}\ \bibnamefont
  {Curiel}}\ and\ \bibinfo {author} {\bibfnamefont {H.~G.}\ \bibnamefont
  {Ramírez}},\ }\href@noop {} {\bibinfo {title} {Vaccination strategies
  against covid-19 and the diffusion of anti-vaccination views}} (\bibinfo
  {year} {2020}),\ \Eprint {https://arxiv.org/abs/2009.13674} {arXiv:2009.13674
  [physics.soc-ph]} \BibitemShut {NoStop}%
\bibitem [{\citenamefont {Johnson}\ \emph
  {et~al.}(2020{\natexlab{b}})\citenamefont {Johnson}, \citenamefont
  {Velásquez}, \citenamefont {Leahy}, \citenamefont {Jha},\ and\ \citenamefont
  {Lupu}}]{johnsonnotsure}%
  \BibitemOpen
  \bibfield  {author} {\bibinfo {author} {\bibfnamefont {N.~F.}\ \bibnamefont
  {Johnson}}, \bibinfo {author} {\bibfnamefont {N.}~\bibnamefont {Velásquez}},
  \bibinfo {author} {\bibfnamefont {R.}~\bibnamefont {Leahy}}, \bibinfo
  {author} {\bibfnamefont {O.}~\bibnamefont {Jha}},\ and\ \bibinfo {author}
  {\bibfnamefont {Y.}~\bibnamefont {Lupu}},\ }\href@noop {} {\bibinfo {title}
  {Not sure? handling hesitancy of covid-19 vaccines}} (\bibinfo {year}
  {2020}{\natexlab{b}}),\ \Eprint {https://arxiv.org/abs/2009.08413}
  {arXiv:2009.08413 [physics.soc-ph]} \BibitemShut {NoStop}%
\bibitem [{\citenamefont {Wang}\ \emph {et~al.}(2016)\citenamefont {Wang},
  \citenamefont {Bauch}, \citenamefont {Bhattacharyya}, \citenamefont
  {d'Onofrio}, \citenamefont {Manfredi}, \citenamefont {Perc}, \citenamefont
  {Perra}, \citenamefont {Salath{\'e}},\ and\ \citenamefont
  {Zhao}}]{wang2016statistical}%
  \BibitemOpen
  \bibfield  {author} {\bibinfo {author} {\bibfnamefont {Z.}~\bibnamefont
  {Wang}}, \bibinfo {author} {\bibfnamefont {C.~T.}\ \bibnamefont {Bauch}},
  \bibinfo {author} {\bibfnamefont {S.}~\bibnamefont {Bhattacharyya}}, \bibinfo
  {author} {\bibfnamefont {A.}~\bibnamefont {d'Onofrio}}, \bibinfo {author}
  {\bibfnamefont {P.}~\bibnamefont {Manfredi}}, \bibinfo {author}
  {\bibfnamefont {M.}~\bibnamefont {Perc}}, \bibinfo {author} {\bibfnamefont
  {N.}~\bibnamefont {Perra}}, \bibinfo {author} {\bibfnamefont
  {M.}~\bibnamefont {Salath{\'e}}},\ and\ \bibinfo {author} {\bibfnamefont
  {D.}~\bibnamefont {Zhao}},\ }\bibfield  {title} {\bibinfo {title}
  {Statistical physics of vaccination},\ }\href
  {https://doi.org/10.1016/j.physrep.2016.10.006} {\bibfield  {journal}
  {\bibinfo  {journal} {Physics Reports}\ }\textbf {\bibinfo {volume} {664}},\
  \bibinfo {pages} {1} (\bibinfo {year} {2016})}\BibitemShut {NoStop}%
\bibitem [{\citenamefont {Salath{\'e}}\ and\ \citenamefont
  {Bonhoeffer}(2008)}]{salathe2008effect}%
  \BibitemOpen
  \bibfield  {author} {\bibinfo {author} {\bibfnamefont {M.}~\bibnamefont
  {Salath{\'e}}}\ and\ \bibinfo {author} {\bibfnamefont {S.}~\bibnamefont
  {Bonhoeffer}},\ }\bibfield  {title} {\bibinfo {title} {The effect of opinion
  clustering on disease outbreaks},\ }\href@noop {} {\bibfield  {journal}
  {\bibinfo  {journal} {Journal of The Royal Society Interface}\ }\textbf
  {\bibinfo {volume} {5}},\ \bibinfo {pages} {1505} (\bibinfo {year}
  {2008})}\BibitemShut {NoStop}%
\bibitem [{\citenamefont {Wu}\ \emph {et~al.}(2016{\natexlab{b}})\citenamefont
  {Wu}, \citenamefont {Ni},\ and\ \citenamefont {Shen}}]{wu2016dynamics}%
  \BibitemOpen
  \bibfield  {author} {\bibinfo {author} {\bibfnamefont {J.}~\bibnamefont
  {Wu}}, \bibinfo {author} {\bibfnamefont {S.}~\bibnamefont {Ni}},\ and\
  \bibinfo {author} {\bibfnamefont {S.}~\bibnamefont {Shen}},\ }\bibfield
  {title} {\bibinfo {title} {Dynamics of public opinion under the influence of
  epidemic spreading},\ }\href@noop {} {\bibfield  {journal} {\bibinfo
  {journal} {International Journal of Modern Physics C}\ }\textbf {\bibinfo
  {volume} {27}},\ \bibinfo {pages} {1650079} (\bibinfo {year}
  {2016}{\natexlab{b}})}\BibitemShut {NoStop}%
\bibitem [{\citenamefont {Coelho}\ and\ \citenamefont
  {Code{\c{c}}o}(2009)}]{coelho2009dynamic}%
  \BibitemOpen
  \bibfield  {author} {\bibinfo {author} {\bibfnamefont {F.~C.}\ \bibnamefont
  {Coelho}}\ and\ \bibinfo {author} {\bibfnamefont {C.~T.}\ \bibnamefont
  {Code{\c{c}}o}},\ }\bibfield  {title} {\bibinfo {title} {Dynamic modeling of
  vaccinating behavior as a function of individual beliefs},\ }\href@noop {}
  {\bibfield  {journal} {\bibinfo  {journal} {PLoS Comput Biol}\ }\textbf
  {\bibinfo {volume} {5}},\ \bibinfo {pages} {e1000425} (\bibinfo {year}
  {2009})}\BibitemShut {NoStop}%
\bibitem [{\citenamefont {Voinson}\ \emph {et~al.}(2015)\citenamefont
  {Voinson}, \citenamefont {Billiard},\ and\ \citenamefont
  {Alvergne}}]{voinson2015beyond}%
  \BibitemOpen
  \bibfield  {author} {\bibinfo {author} {\bibfnamefont {M.}~\bibnamefont
  {Voinson}}, \bibinfo {author} {\bibfnamefont {S.}~\bibnamefont {Billiard}},\
  and\ \bibinfo {author} {\bibfnamefont {A.}~\bibnamefont {Alvergne}},\
  }\bibfield  {title} {\bibinfo {title} {Beyond rational decision-making:
  modelling the influence of cognitive biases on the dynamics of vaccination
  coverage},\ }\href@noop {} {\bibfield  {journal} {\bibinfo  {journal} {PloS
  one}\ }\textbf {\bibinfo {volume} {10}},\ \bibinfo {pages} {e0142990}
  (\bibinfo {year} {2015})}\BibitemShut {NoStop}%
\bibitem [{\citenamefont {Fu}\ \emph {et~al.}(2017)\citenamefont {Fu},
  \citenamefont {Christakis},\ and\ \citenamefont {Fowler}}]{Feng2017}%
  \BibitemOpen
  \bibfield  {author} {\bibinfo {author} {\bibfnamefont {F.}~\bibnamefont
  {Fu}}, \bibinfo {author} {\bibfnamefont {N.~A.}\ \bibnamefont {Christakis}},\
  and\ \bibinfo {author} {\bibfnamefont {J.~H.}\ \bibnamefont {Fowler}},\
  }\bibfield  {title} {\bibinfo {title} {Dueling biological and social
  contagions},\ }\href {https://doi.org/10.1038/srep43634} {\bibfield
  {journal} {\bibinfo  {journal} {Scientific Reports}\ }\textbf {\bibinfo
  {volume} {7}},\ \bibinfo {pages} {43634} (\bibinfo {year}
  {2017})}\BibitemShut {NoStop}%
\bibitem [{\citenamefont {Lallouache}\ \emph {et~al.}(2010)\citenamefont
  {Lallouache}, \citenamefont {Chakrabarti}, \citenamefont {Chakraborti},\ and\
  \citenamefont {Chakrabarti}}]{2010lallouacheCCC}%
  \BibitemOpen
  \bibfield  {author} {\bibinfo {author} {\bibfnamefont {M.}~\bibnamefont
  {Lallouache}}, \bibinfo {author} {\bibfnamefont {A.~S.}\ \bibnamefont
  {Chakrabarti}}, \bibinfo {author} {\bibfnamefont {A.}~\bibnamefont
  {Chakraborti}},\ and\ \bibinfo {author} {\bibfnamefont {B.~K.}\ \bibnamefont
  {Chakrabarti}},\ }\bibfield  {title} {\bibinfo {title} {Opinion formation in
  kinetic exchange models: Spontaneous symmetry-breaking transition},\ }\href
  {https://doi.org/10.1103/PhysRevE.82.056112} {\bibfield  {journal} {\bibinfo
  {journal} {Phys. Rev. E}\ }\textbf {\bibinfo {volume} {82}},\ \bibinfo
  {pages} {056112} (\bibinfo {year} {2010})}\BibitemShut {NoStop}%
\bibitem [{\citenamefont {Zeng}\ and\ \citenamefont
  {Chen}(2005)}]{zeng2005complexity}%
  \BibitemOpen
  \bibfield  {author} {\bibinfo {author} {\bibfnamefont {G.-Z.}\ \bibnamefont
  {Zeng}}\ and\ \bibinfo {author} {\bibfnamefont {L.-S.}\ \bibnamefont
  {Chen}},\ }\bibfield  {title} {\bibinfo {title} {Complexity and asymptotical
  behavior of a sirs epidemic model with proportional impulsive vaccination},\
  }\href@noop {} {\bibfield  {journal} {\bibinfo  {journal} {Advances in
  Complex Systems}\ }\textbf {\bibinfo {volume} {8}},\ \bibinfo {pages} {419}
  (\bibinfo {year} {2005})}\BibitemShut {NoStop}%
\bibitem [{\citenamefont {Rao}\ \emph {et~al.}(2019)\citenamefont {Rao},
  \citenamefont {Mandal},\ and\ \citenamefont {Kang}}]{rao2019complicated}%
  \BibitemOpen
  \bibfield  {author} {\bibinfo {author} {\bibfnamefont {F.}~\bibnamefont
  {Rao}}, \bibinfo {author} {\bibfnamefont {P.~S.}\ \bibnamefont {Mandal}},\
  and\ \bibinfo {author} {\bibfnamefont {Y.}~\bibnamefont {Kang}},\ }\bibfield
  {title} {\bibinfo {title} {Complicated endemics of an sirs model with a
  generalized incidence under preventive vaccination and treatment controls},\
  }\href@noop {} {\bibfield  {journal} {\bibinfo  {journal} {Applied
  Mathematical Modelling}\ }\textbf {\bibinfo {volume} {67}},\ \bibinfo {pages}
  {38} (\bibinfo {year} {2019})}\BibitemShut {NoStop}%
\bibitem [{\citenamefont {Moneim}\ and\ \citenamefont
  {Greenhalgh}(2005)}]{moneim2005threshold}%
  \BibitemOpen
  \bibfield  {author} {\bibinfo {author} {\bibfnamefont {I.}~\bibnamefont
  {Moneim}}\ and\ \bibinfo {author} {\bibfnamefont {D.}~\bibnamefont
  {Greenhalgh}},\ }\bibfield  {title} {\bibinfo {title} {Threshold and
  stability results for an sirs epidemic model with a general periodic
  vaccination strategy},\ }\href@noop {} {\bibfield  {journal} {\bibinfo
  {journal} {Journal of biological systems}\ }\textbf {\bibinfo {volume}
  {13}},\ \bibinfo {pages} {131} (\bibinfo {year} {2005})}\BibitemShut
  {NoStop}%
\bibitem [{\citenamefont {Lahrouz}\ \emph {et~al.}(2012)\citenamefont
  {Lahrouz}, \citenamefont {Omari}, \citenamefont {Kiouach},\ and\
  \citenamefont {Belma{\^a}ti}}]{lahrouz2012complete}%
  \BibitemOpen
  \bibfield  {author} {\bibinfo {author} {\bibfnamefont {A.}~\bibnamefont
  {Lahrouz}}, \bibinfo {author} {\bibfnamefont {L.}~\bibnamefont {Omari}},
  \bibinfo {author} {\bibfnamefont {D.}~\bibnamefont {Kiouach}},\ and\ \bibinfo
  {author} {\bibfnamefont {A.}~\bibnamefont {Belma{\^a}ti}},\ }\bibfield
  {title} {\bibinfo {title} {Complete global stability for an sirs epidemic
  model with generalized non-linear incidence and vaccination},\ }\href@noop {}
  {\bibfield  {journal} {\bibinfo  {journal} {Applied Mathematics and
  Computation}\ }\textbf {\bibinfo {volume} {218}},\ \bibinfo {pages} {6519}
  (\bibinfo {year} {2012})}\BibitemShut {NoStop}%
\bibitem [{\citenamefont {Doutor}\ \emph {et~al.}(2016)\citenamefont {Doutor},
  \citenamefont {Rodrigues}, \citenamefont {do~C{\'e}u~Soares},\ and\
  \citenamefont {Chalub}}]{doutor2016optimal}%
  \BibitemOpen
  \bibfield  {author} {\bibinfo {author} {\bibfnamefont {P.}~\bibnamefont
  {Doutor}}, \bibinfo {author} {\bibfnamefont {P.}~\bibnamefont {Rodrigues}},
  \bibinfo {author} {\bibfnamefont {M.}~\bibnamefont {do~C{\'e}u~Soares}},\
  and\ \bibinfo {author} {\bibfnamefont {F.~A.}\ \bibnamefont {Chalub}},\
  }\bibfield  {title} {\bibinfo {title} {Optimal vaccination strategies and
  rational behaviour in seasonal epidemics},\ }\href@noop {} {\bibfield
  {journal} {\bibinfo  {journal} {Journal of mathematical biology}\ }\textbf
  {\bibinfo {volume} {73}},\ \bibinfo {pages} {1437} (\bibinfo {year}
  {2016})}\BibitemShut {NoStop}%
\bibitem [{\citenamefont {Vel{\'a}squez-Rojas}\ and\ \citenamefont
  {Vazquez}(2017)}]{velasquez2017interacting}%
  \BibitemOpen
  \bibfield  {author} {\bibinfo {author} {\bibfnamefont {F.}~\bibnamefont
  {Vel{\'a}squez-Rojas}}\ and\ \bibinfo {author} {\bibfnamefont
  {F.}~\bibnamefont {Vazquez}},\ }\bibfield  {title} {\bibinfo {title}
  {Interacting opinion and disease dynamics in multiplex networks:
  discontinuous phase transition and nonmonotonic consensus times},\
  }\href@noop {} {\bibfield  {journal} {\bibinfo  {journal} {Physical Review
  E}\ }\textbf {\bibinfo {volume} {95}},\ \bibinfo {pages} {052315} (\bibinfo
  {year} {2017})}\BibitemShut {NoStop}%
\bibitem [{\citenamefont {Ventura}\ \emph {et~al.}(2021)\citenamefont
  {Ventura}, \citenamefont {Moreno},\ and\ \citenamefont
  {Rodrigues}}]{ventura2021role}%
  \BibitemOpen
  \bibfield  {author} {\bibinfo {author} {\bibfnamefont {P.~C.}\ \bibnamefont
  {Ventura}}, \bibinfo {author} {\bibfnamefont {Y.}~\bibnamefont {Moreno}},\
  and\ \bibinfo {author} {\bibfnamefont {F.~A.}\ \bibnamefont {Rodrigues}},\
  }\bibfield  {title} {\bibinfo {title} {Role of time scale in the spreading of
  asymmetrically interacting diseases},\ }\href@noop {} {\bibfield  {journal}
  {\bibinfo  {journal} {Physical Review Research}\ }\textbf {\bibinfo {volume}
  {3}},\ \bibinfo {pages} {013146} (\bibinfo {year} {2021})}\BibitemShut
  {NoStop}%
\bibitem [{\citenamefont {da~Silva}\ \emph {et~al.}(2019)\citenamefont
  {da~Silva}, \citenamefont {Vel{\'a}squez-Rojas}, \citenamefont {Connaughton},
  \citenamefont {Vazquez}, \citenamefont {Moreno},\ and\ \citenamefont
  {Rodrigues}}]{da2019epidemic}%
  \BibitemOpen
  \bibfield  {author} {\bibinfo {author} {\bibfnamefont {P.~C.~V.}\
  \bibnamefont {da~Silva}}, \bibinfo {author} {\bibfnamefont {F.}~\bibnamefont
  {Vel{\'a}squez-Rojas}}, \bibinfo {author} {\bibfnamefont {C.}~\bibnamefont
  {Connaughton}}, \bibinfo {author} {\bibfnamefont {F.}~\bibnamefont
  {Vazquez}}, \bibinfo {author} {\bibfnamefont {Y.}~\bibnamefont {Moreno}},\
  and\ \bibinfo {author} {\bibfnamefont {F.~A.}\ \bibnamefont {Rodrigues}},\
  }\bibfield  {title} {\bibinfo {title} {Epidemic spreading with awareness and
  different timescales in multiplex networks},\ }\href@noop {} {\bibfield
  {journal} {\bibinfo  {journal} {Physical Review E}\ }\textbf {\bibinfo
  {volume} {100}},\ \bibinfo {pages} {032313} (\bibinfo {year}
  {2019})}\BibitemShut {NoStop}%
\bibitem [{\citenamefont {Oestereich}\ \emph {et~al.}(2019)\citenamefont
  {Oestereich}, \citenamefont {Pires},\ and\ \citenamefont
  {Crokidakis}}]{2019oestereichPC}%
  \BibitemOpen
  \bibfield  {author} {\bibinfo {author} {\bibfnamefont {A.~L.}\ \bibnamefont
  {Oestereich}}, \bibinfo {author} {\bibfnamefont {M.~A.}\ \bibnamefont
  {Pires}},\ and\ \bibinfo {author} {\bibfnamefont {N.}~\bibnamefont
  {Crokidakis}},\ }\bibfield  {title} {\bibinfo {title} {Three-state opinion
  dynamics in modular networks},\ }\href
  {https://doi.org/10.1103/physreve.100.032312} {\bibfield  {journal} {\bibinfo
   {journal} {Phys. Rev. E}\ }\textbf {\bibinfo {volume} {100}},\ \bibinfo
  {pages} {032312} (\bibinfo {year} {2019})}\BibitemShut {NoStop}%
\bibitem [{\citenamefont {Wang}\ \emph {et~al.}(2020)\citenamefont {Wang},
  \citenamefont {Sirianni}, \citenamefont {Tang}, \citenamefont {Zheng},\ and\
  \citenamefont {Fu}}]{Wang2020}%
  \BibitemOpen
  \bibfield  {author} {\bibinfo {author} {\bibfnamefont {X.}~\bibnamefont
  {Wang}}, \bibinfo {author} {\bibfnamefont {A.~D.}\ \bibnamefont {Sirianni}},
  \bibinfo {author} {\bibfnamefont {S.}~\bibnamefont {Tang}}, \bibinfo {author}
  {\bibfnamefont {Z.}~\bibnamefont {Zheng}},\ and\ \bibinfo {author}
  {\bibfnamefont {F.}~\bibnamefont {Fu}},\ }\bibfield  {title} {\bibinfo
  {title} {Public discourse and social network echo chambers driven by
  socio-cognitive biases},\ }\href {https://doi.org/10.1103/PhysRevX.10.041042}
  {\bibfield  {journal} {\bibinfo  {journal} {Phys. Rev. X}\ }\textbf {\bibinfo
  {volume} {10}},\ \bibinfo {pages} {041042} (\bibinfo {year}
  {2020})}\BibitemShut {NoStop}%
\bibitem [{\citenamefont {Kadelka}\ and\ \citenamefont
  {McCombs}(2020)}]{kadelka2020effect}%
  \BibitemOpen
  \bibfield  {author} {\bibinfo {author} {\bibfnamefont {C.}~\bibnamefont
  {Kadelka}}\ and\ \bibinfo {author} {\bibfnamefont {A.}~\bibnamefont
  {McCombs}},\ }\bibfield  {title} {\bibinfo {title} {Effect of clustering and
  correlation of belief systems on infectious disease outbreaks},\ }\href@noop
  {} {\bibfield  {journal} {\bibinfo  {journal} {medRxiv}\ } (\bibinfo {year}
  {2020})}\BibitemShut {NoStop}%
\bibitem [{\citenamefont {Bizzarri}\ \emph {et~al.}(2021)\citenamefont
  {Bizzarri}, \citenamefont {Panebianco},\ and\ \citenamefont
  {Pin}}]{bizzarri2021epidemic}%
  \BibitemOpen
  \bibfield  {author} {\bibinfo {author} {\bibfnamefont {M.}~\bibnamefont
  {Bizzarri}}, \bibinfo {author} {\bibfnamefont {F.}~\bibnamefont
  {Panebianco}},\ and\ \bibinfo {author} {\bibfnamefont {P.}~\bibnamefont
  {Pin}},\ }\href@noop {} {\bibinfo {title} {Epidemic dynamics with homophily,
  vaccination choices, and pseudoscience attitudes}} (\bibinfo {year} {2021}),\
  \Eprint {https://arxiv.org/abs/2007.08523} {arXiv:2007.08523 [q-bio.PE]}
  \BibitemShut {NoStop}%
\bibitem [{\citenamefont {Saad-Roy}\ \emph {et~al.}(2020)\citenamefont
  {Saad-Roy}, \citenamefont {Wagner}, \citenamefont {Baker}, \citenamefont
  {Morris}, \citenamefont {Farrar}, \citenamefont {Graham}, \citenamefont
  {Levin}, \citenamefont {Mina}, \citenamefont {Metcalf},\ and\ \citenamefont
  {Grenfell}}]{saad2020immune}%
  \BibitemOpen
  \bibfield  {author} {\bibinfo {author} {\bibfnamefont {C.~M.}\ \bibnamefont
  {Saad-Roy}}, \bibinfo {author} {\bibfnamefont {C.~E.}\ \bibnamefont
  {Wagner}}, \bibinfo {author} {\bibfnamefont {R.~E.}\ \bibnamefont {Baker}},
  \bibinfo {author} {\bibfnamefont {S.~E.}\ \bibnamefont {Morris}}, \bibinfo
  {author} {\bibfnamefont {J.}~\bibnamefont {Farrar}}, \bibinfo {author}
  {\bibfnamefont {A.~L.}\ \bibnamefont {Graham}}, \bibinfo {author}
  {\bibfnamefont {S.~A.}\ \bibnamefont {Levin}}, \bibinfo {author}
  {\bibfnamefont {M.~J.}\ \bibnamefont {Mina}}, \bibinfo {author}
  {\bibfnamefont {C.~J.~E.}\ \bibnamefont {Metcalf}},\ and\ \bibinfo {author}
  {\bibfnamefont {B.~T.}\ \bibnamefont {Grenfell}},\ }\bibfield  {title}
  {\bibinfo {title} {Immune life history, vaccination, and the dynamics of
  sars-cov-2 over the next 5 years},\ }\href@noop {} {\bibfield  {journal}
  {\bibinfo  {journal} {Science}\ }\textbf {\bibinfo {volume} {370}},\ \bibinfo
  {pages} {811} (\bibinfo {year} {2020})}\BibitemShut {NoStop}%
\bibitem [{\citenamefont {Marks}\ and\ \citenamefont
  {Vanderslott}(2021)}]{ourworldindata}%
  \BibitemOpen
  \bibfield  {author} {\bibinfo {author} {\bibfnamefont {T.}~\bibnamefont
  {Marks}}\ and\ \bibinfo {author} {\bibfnamefont {S.}~\bibnamefont
  {Vanderslott}},\ }\bibfield  {title} {\bibinfo {title} {Which countries have
  mandatory childhood vaccination policies?},\ }\href
  {https://ourworldindata.org/childhood-vaccination-policies} {\bibfield
  {journal} {\bibinfo  {journal} {Our World in Data}\ } (\bibinfo {year}
  {2021})}\BibitemShut {NoStop}%
\bibitem [{\citenamefont {Oestereich}\ \emph {et~al.}(2020)\citenamefont
  {Oestereich}, \citenamefont {Pires}, \citenamefont {Queir{\'o}s},\ and\
  \citenamefont {Crokidakis}}]{oestereich2020hysteresis}%
  \BibitemOpen
  \bibfield  {author} {\bibinfo {author} {\bibfnamefont {A.~L.}\ \bibnamefont
  {Oestereich}}, \bibinfo {author} {\bibfnamefont {M.~A.}\ \bibnamefont
  {Pires}}, \bibinfo {author} {\bibfnamefont {S.~D.}\ \bibnamefont
  {Queir{\'o}s}},\ and\ \bibinfo {author} {\bibfnamefont {N.}~\bibnamefont
  {Crokidakis}},\ }\bibfield  {title} {\bibinfo {title} {Hysteresis and
  disorder-induced order in continuous kinetic-like opinion dynamics in complex
  networks},\ }\href@noop {} {\bibfield  {journal} {\bibinfo  {journal} {Chaos,
  Solitons \& Fractals}\ }\textbf {\bibinfo {volume} {137}},\ \bibinfo {pages}
  {109893} (\bibinfo {year} {2020})}\BibitemShut {NoStop}%
\end{thebibliography}%

\end{document}